\documentclass[ ]{iopart}
\usepackage{graphicx}
\usepackage{amsfonts}
\usepackage{amssymb}
\usepackage{graphicx}
\usepackage[english]{babel}
\usepackage[utf8]{inputenc}
\usepackage{units}
\usepackage[arrow, matrix, curve]{xy}
\newcommand{\tu}[1]{\textup{#1}}

\newcommand{\Abb}[4]{\left\{ \begin{array}{ccc}
                               #1 & \rightarrow &#2\\
			       #3 &\mapsto &#4
                               \end{array}\right.}

\newcommand{\N}{\mathbb{N}}
\newcommand{\R}{\mathbb R}
\newcommand{\Z}{\mathbb Z}
\newcommand{\C}{\mathbb{C}}

\begin{document}
\title{Resonance chains in open systems, generalized zeta functions and clustering of the length spectrum}

\author{S.~Barkhofen$^{3,2}$, F.~Faure$^{4}$, T.~Weich$^{1,2,*}$}
\address{$^1$ Fachbereich Mathematik, Philipps-Universit\"{a}t Marburg, Hans-Meerwein-Stra{\ss}e,
35032 Marburg, Germany}
\address{$^2$Fachbereich Physik, Philipps-Universit\"{a}t Marburg, Renthof 5,
35032 Marburg, Germany}
\address{$^3${Applied Physics, University of Paderborn, Warburger Strasse 100, 33098 Paderborn, Germany}}
\address{$^4$Institut Fourier,  100 rue des maths, BP 74, 38402 St Martin d'H\`eres cedex, France}
\ead{*: weich@mathematik.uni-marburg.de}

\date{\today}
\begin{abstract}
In many non-integrable open systems in physics and mathematics resonances have
been found to be surprisingly ordered along curved lines in 
the complex plane. In this article we provide a unifying approach to 
these resonance chains by generalizing dynamical zeta functions.
By means of a detailed numerical study we show that these generalized zeta 
functions explain the mechanism that creates the chains 
of quantum resonance and classical Ruelle resonances for 
3-disk systems as well as geometric resonances on Schottky surfaces.
We also present a direct system-intrinsic definition of the 
continuous lines on which the resonances are strung together as a 
projection of an analytic variety. Additionally, this approach shows that
the existence of resonance chains is directly related to a clustering 
of the classical length spectrum on multiples of a base length. Finally, this 
link is used to construct new examples where several 
different structures of resonance chains coexist. 
\end{abstract}


\maketitle

\section{Introduction}
\label{sec:Introduction}
The study of resonances in scattering systems with chaotic classical dynamics
is an active field of research. It is at the same time driven by physicists being 
interested in the quantum properties of open chaotic systems as well 
as by mathematicians being interested in geometric analysis and partial differential equations (PDE)
on non-compact manifolds. While in scattering theory, resonances
are poles of the scattering matrix and correspond to quasi-stationary 
quantum states with purely in- or outgoing boundary conditions, from the PDE  
point of view resonances are often seen as the spectral invariants defined as the poles of
a meromorphic continuation of the resolvent operator. 

Aside from some very special
exceptions (e.g. the Walsh quantized Baker map \cite{non07b}) the position of 
the resonances of such complex systems cannot be determined exactly. 
It has, however, been a very successful approach to gain coarser 
information on general properties of the resonance distribution
in terms of the underlying classical dynamics and geometry, respectively. For 
example results on spectral gaps \cite{ika88,non09,pet10} predict resonance free regions and 
the fractal Weyl law \cite{sjo90,non14, gui04,lu03,sch04e} gives an upper bound on the growth of the 
number of the so-called ``long-living'' resonances in the high frequency 
limit. Despite the large progress that has been made in obtaining rigorous 
results on the distribution of the resonances, there are still many 
open questions (we refer to \cite{non11} for an overview over recent results 
and open questions). For example the 
upper bound on the exponential growth of the resonance density which has 
been proven in the fractal Weyl law \cite{non14, gui04, sjo90} is only 
conjectured to be sharp and there
exists also a conjecture on the improvement of the spectral gap estimates
\cite{jak12}. It
has thus been a very important approach to study the resonance structure 
numerically and by physical experiments in order to test the present 
conjectures and to show that the results on resonance distributions can 
be observed in real physical systems. There have been established two 
paradigmatic scattering systems for these kind of tests: the  
$n$-disk systems and Schottky surfaces. 

A $n$-disk system describes the scattering of one particle 
at $n$-circular obstacles in the Euclidean plane (see \fref{fig:3-disk}). 
It has been introduced
by Gaspard-Rice \cite{gas89b,gas89a,gas89d} and Cvitanovi\'c-Eckhardt \cite{cvi89} 
in physics and  by Ikawa \cite{ika88} in mathematics. In physics
especially the completely symmetric 3-disk system, where 3 disks of the same size
are arranged on a regular triangle, enjoys great popularity. It is 
already complex enough to have a classical dynamic with a non-trivial, fractal 
repeller and is still simple enough to be treated efficiently by numerics. There exist 
very efficient approaches to calculate the quantum resonances based on
a direct scattering matrix ansatz \cite{gas89d} as well as by a semiclassical approach 
using zeta functions \cite{cvi89}. Consequently it has been among the first systems
to numerically observe the fractal Weyl law \cite{lu03}. Further 
importance arises from its experimental realizability \cite{lu99} and 
in recent microwave experiments the spectral gap \cite{bar13b} and first hints of the 
fractal Weyl law \cite{pot12} have been found. From a mathematical point of view 
the disadvantage of the $n$-disk system is that many of the numerical algorithms 
are not based on mathematical rigorous results. For example the conjecture that
the zeros of the semiclassical zeta function are related to the quantum 
resonances, which has been thoroughly tested in physics \cite{cvi89,wir99b}, is mathematically 
open (cf. \cite[Section 1.6-1.8]{arXfau12} for recent results into this direction).

This is the reason for the popularity of Schottky surfaces as an explicit model 
system in mathematics. Schottky surfaces are two-dimensional non-compact surfaces with
constant negative curvature which are convex co-compact (see \fref{fig:schottky} for a sketch
of a 3-funneled Schottky surface).  
The constant negative curvature allows to prove that
the zeros of the Selberg zeta function are related to the resonances
of the Laplacian \cite{pat01} and not only approximately as conjectured  
for the 3-disk system in physics. Furthermore
Jenkinson-Pollicott \cite{jen02} derived,
analogously to the cycle expansion of the 3-disk system,
efficient formulas to numerically compute
the zeros of the Selberg zeta functions. The big drawback of these surfaces from the 
physical point of view is that there is no known experimental 
realization (see \cite{sch96,sto07} for attempts to realize the wave equation on negatively curved surfaces 
with water waves). As the curvature of these surfaces is, however, strictly negative, 
the geodesic flow is hyperbolic and these surfaces are an interesting model 
for open chaotic systems. Additionally the study of the resonances on constant negative curvature 
surfaces and the relation to their geometry is a mathematically interesting question itself 
(see e.g. \cite{zwo99}) thus we will call these resonances \emph{geometric resonances}. 
The most simple example of Schottky surfaces which already 
has a non-trivial fractal repeller are the 3-funneled surfaces 
as shown in the upper part of \fref{fig:schottky}. For these surfaces Borthwick recently presented a thorough numerical study 
of their resonances structure \cite{bor14}. 

\begin{figure}
\centering
        \includegraphics[width=0.8\textwidth]{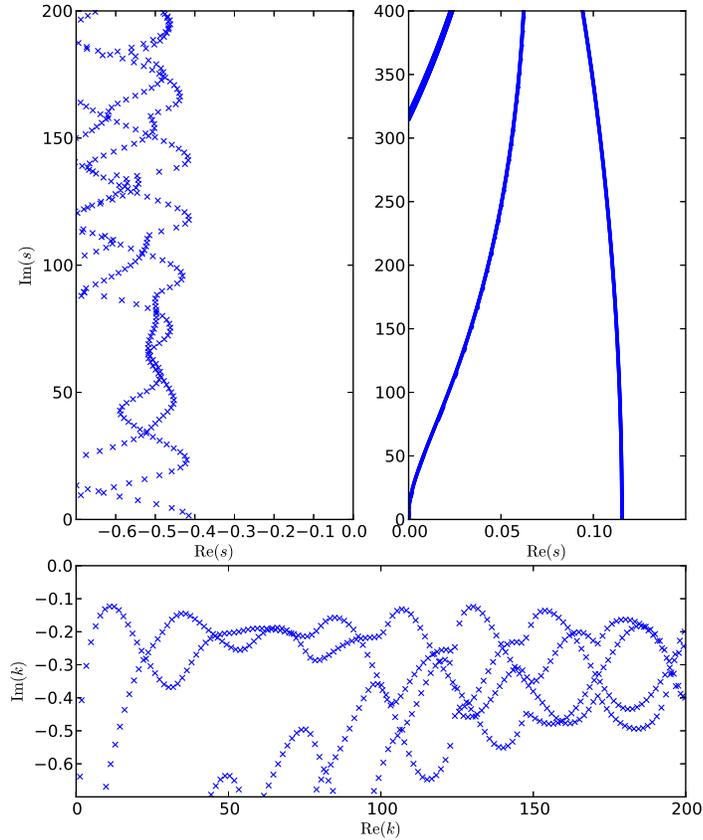}
\caption{Resonance structure of the classical Ruelle resonances
of the 3-disk system (upper left), the Schottky surface $X_{12,12,12}$ (upper right) and
the quantum 3-disk system with $R/a=6$ (lower). Note: the resonances for the Schottky surfaces
are so dense that they appear as continuous lines in the plot above, however, they are discrete
(see \fref{fig:rot_res_BS}).
}\label{fig:res_chains}
\end{figure}

One of the most surprising features of this resonance structure is an observation which 
has before already been made for the 3-disk system \cite{gas89d, wir99b}: Even though the 
underlying classical dynamic is completely hyperbolic, i.e.~strongly chaotic, 
the distribution of the resonances is highly ordered and the resonances form a striking 
chain structure in the complex plane (see \fref{fig:res_chains}). The 3-disk system and the Schottky 
surfaces are, however, not the only systems to embody these chain structures. Very similar 
chains have for example also been observed in higher dimensional systems, as the 4-sphere 
billiard \cite{ebe10}, and even in microdisk resonators which are important models for physical 
applications such as microdisk lasers \cite{wie08}. Furthermore the phenomenon of resonance chains is
not only restricted to resonances of the Laplace operator and quantum resonances, 
respectively. Also the spectrum of the classical Ruelle resonances which describe the decay 
of correlations in classical dynamical systems has been reported to reveal these clear 
resonance chains \cite{gas92}. 

Though these chains are eye-catching, little is known about these
structures. For the completely symmetric 3-disk system and the completely symmetric
3-funneled Schottky surface, the distance between the resonances on one chain has been 
observed to be related to the length of the shortest periodic orbit \cite{bor14, wir99b}. 
Additionally the oscillation length of the chains has been found to be
related to the length difference between two fundamental periodic orbits. 
Recently a numerical study of the 3-disk system \cite{wei14}
showed that the resonance chains are not only a visual effect, but that the 
resonances on one chain share common physical properties. It was, however, shown that the resonance chains are not caused by scarring of the scattering states along individual classical orbits, but are related to the interatction of several classical orbits. 

Two fundamental questions are, however, still open and will 
be addressed in this article. First of all the resonance chains, are a visual 
impression, i.e.~our brain groups the resonances to different chains and we have 
the impression that they are strung along different continuous lines. In this
article we will give a system intrinsic definition of these continuous lines. 
Secondly, resonance chains have been observed in many, quite different systems. 
We will give a unifying condition on the classical length spectrum that is able 
to explain and predict the existence of resonance chains.

The article is organized as follows: In Section~\ref{sec:3-disk} we treat the 
resonance chains of the symmetric 3-disk system. After a general 
introduction of 3-disk systems
(Section~\ref{sec:intro_3-disk}) we will numerically illustrate 
that the resonance chains of the symmetric 3-disk system are created 
by fast rotating eigenvalues of a suitable transfer operator on the canonical 
Poincar\'e section. This idea applies for the semiclassical
as well as for the classical resonances (Section~\ref{sec:rot_res}). 
Then we give an introduction
to Schottky surfaces (Section~\ref{sec:intro_Schottky}) and show in 
Section~\ref{sec:gen_zeta} that a straightforward application 
of this idea on the resonance spectrum of completely symmetric Schottky surfaces
fails to explain the resonance chains. However, the ideas can be abstracted. This 
leads us to the introduction of a generalized zeta function and the generalized spectrum
which are shown to be the right objects to describe and understand the resonance chains. 
These objects immediately also allow to deal with non-symmetric Schottky
surfaces and 3-disk systems and also gives a condition for the existence of 
resonance chains (Section~\ref{sec:existence_of_chains}). In 
Section~\ref{sec:num_test}  we will finally test this condition in different 
cases and use its predictive power to identify systems incorporating different
coexistent chain structures. 

As this article addresses a physical as well as a mathematical readership we decided
to always state clearly which results and relations are known to be true by 
mathematically rigorous proofs and which results are achieved by physical 
arguments or experiments. This will help to keep an overview which parts of the argumentation
are based on rigorous arguments and where further mathematical work remains to be done. 

This article itself does not claim to derive any rigorous statements either. 
Instead our aim is to gain a universal understanding of the appearance of the
resonance chains and to identify the mathematical quantities,
which are crucial for their understanding, via numerical investigations. As the question, whether a chain 
structure is visible or not, is mathematically not well posed but depends on the
observer of the resonance structure, our condition cannot be a \emph{if and only if} 
condition either. The condition established in Section~\ref{sec:gen_zeta} is thus an approximate
condition and our hypothesis is that the better this condition is fulfilled the 
clearer the chains will be visible. This hypothesis is tested and supported by various
numerical results presented in Section~\ref{sec:num_test}. A rigorouse proof of the 
existence of resonance will be given in \cite{mathArticle}, where the ideas and insights
presented in this article are an essential ingredient.

\section{3-disk systems and rotating resonances}\label{sec:3-disk}
\subsection{Introduction to 3-disk systems}\label{sec:intro_3-disk}
\begin{figure}
\centering
        \includegraphics[width=0.6\textwidth]{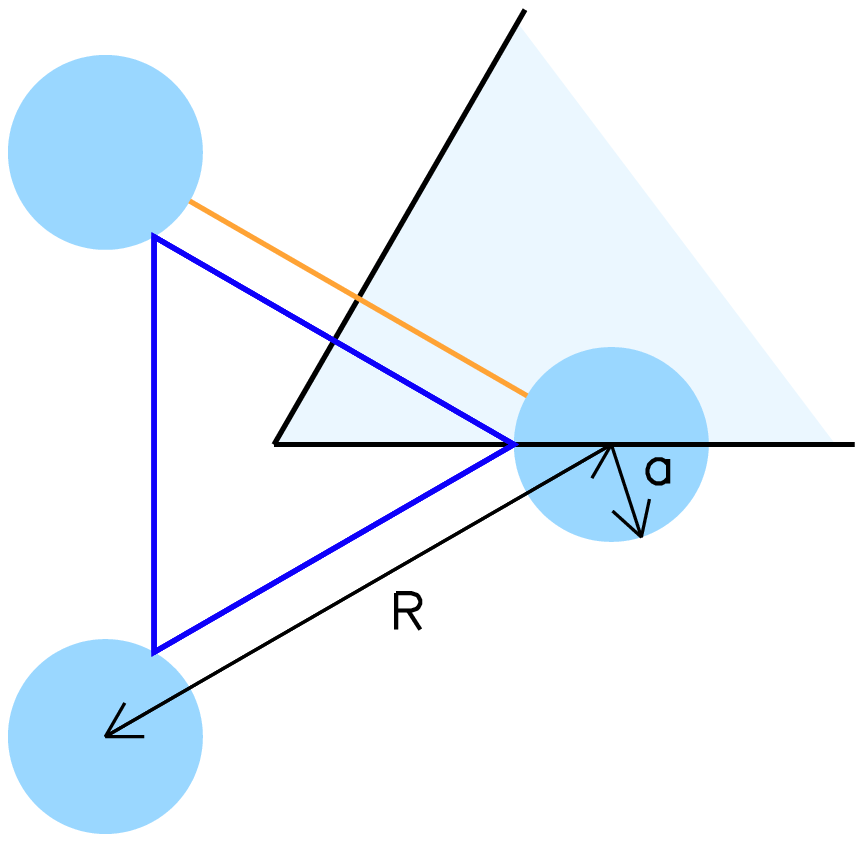}
        \includegraphics[width=0.39\textwidth]{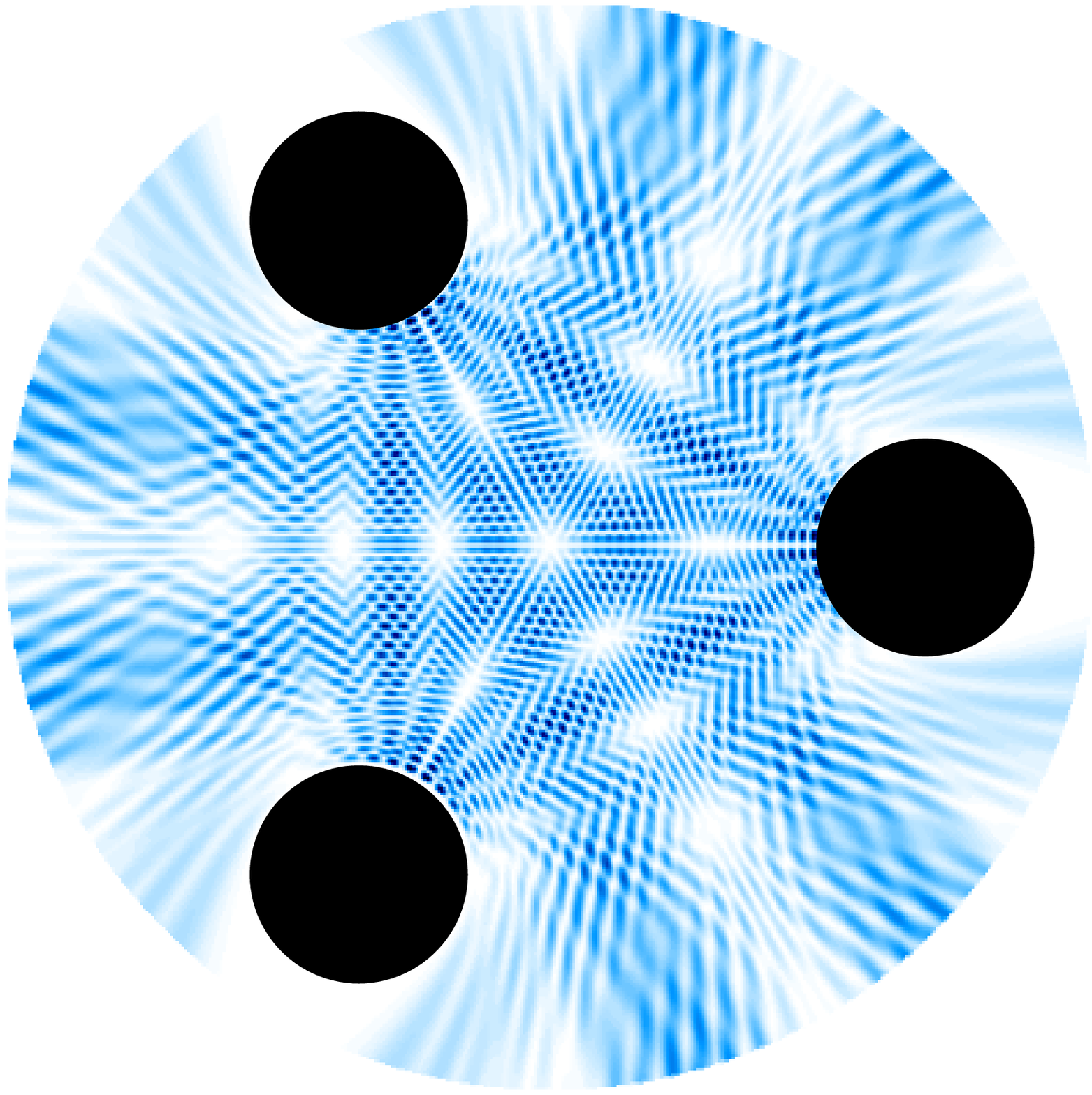}
\caption{Left: Sketch of a symmetric 3-disk system and two classical 
closed trajectories (yellow and blue lines). The fundamental domain is shaded
in light blue. The symmetry reduced billiard is then a billiard consisting
of the fundamental domain with two additional hard walls along the symmetry axes. 
Right: Quantum mechanical scattering state
with Dirichlet boundary conditions and purely outgoing boundary conditions 
calculated by the algorithm as described in \cite{wei14}. Note that this
scattering state is additionally vanishing along the symmetry 
axes and is thus also a scattering state of the symmetry reduced billiard
with Dirichlet boundary conditions.
}
\label{fig:3-disk}
\end{figure}

A 3-disk system describes one free particle in an Euclidean 2-dimensional space that
scatters at three non-overlapping hard disks $D_1, D_2$ and $D_3$. For further use 
we will call the union of the disks by $D=\bigcup_{i} D_i$. Classically 
its trajectories are given by straight lines and hard wall reflections at the 
disk boundaries $\partial D$ (see left panel of \fref{fig:3-disk}). Its 
quantum dynamics is described by the Laplacian $\Delta$ with Dirichlet boundary conditions
at the disk boundaries and its quantum resonances can rigorously be defined
as the poles of the resolvent
\begin{equation}\label{eq:resolvent_3_disk}
 R(k):=(-\Delta -k^2)^{-1}:
    L^2_{\mathrm{comp}}(\R^2\setminus D)\to L^2_{\mathrm{loc}}(\R^2\setminus D).
\end{equation}
This resolvent, as a map between compactly supported $L^2$ functions and locally 
$L^2$-integrable functions, is known to extend meromorphically to the $\mathbb C \setminus l$ where 
$l$ is any line connecting zero with infinity \cite{vai68}. 
These poles do indeed coincide with the physical 
scattering resonances i.e.~if $k_n$ is a pole with multiplicity $m$, 
then there exist $m$ solutions of 
\[
 (-\Delta - k_n^2)\Psi=0
\]
with Dirichlet and purely outgoing boundary conditions. The latter condition
means that the solution $\Psi(r,\phi)$ in polar coordinates $(r,\phi)$ can be 
written asymptotically in the limit $r\to\infty$ as \cite{gas89d}
\[
 \Psi(r,\phi) \sim \sum\limits_{l=-\infty}^\infty C_l\frac{\exp(i(kr-l\pi/2-\pi/4))}{\sqrt{2\pi k r}} \exp(il\phi).
\]
Contrary purely incoming boundary conditions would correspond to 
\[
 \Psi(r,\phi) \sim \sum\limits_{l=-\infty}^\infty C_l\frac{\exp(-i(kr-l\pi/2-\pi/4))}{\sqrt{2\pi k r}} \exp(il\phi).
\]
Here $C_l$ are constant coefficients and for fixed $l$ the asymptotic $r,\phi$
dependence equals the asymptotic dependence of a free particle state with angular
momentum $l$. 

An important class of 3-disk systems are those which are completely symmetric. 
In those systems all three disks have the same size and the disks are arranged 
on an equilateral triangle (see left part of \fref{fig:3-disk}). Historically the side length of the triangle is 
denoted by $R$ and the disk radius by $a$ \cite{gas89a}. These systems are then
completely characterized, up to a scaling, by the ratio $R/a$. As the system
is highly symmetric and possesses the symmetry group $D_3$ of the equilateral triangle 
often the symmetry reduced system is studied in physics.
This system consists of an open billiard representing the fundamental domain
of the 3-disk system (see light shaded region in the left part of \fref{fig:3-disk}).

\begin{figure}
\centering
        \includegraphics[width=0.7\textwidth]{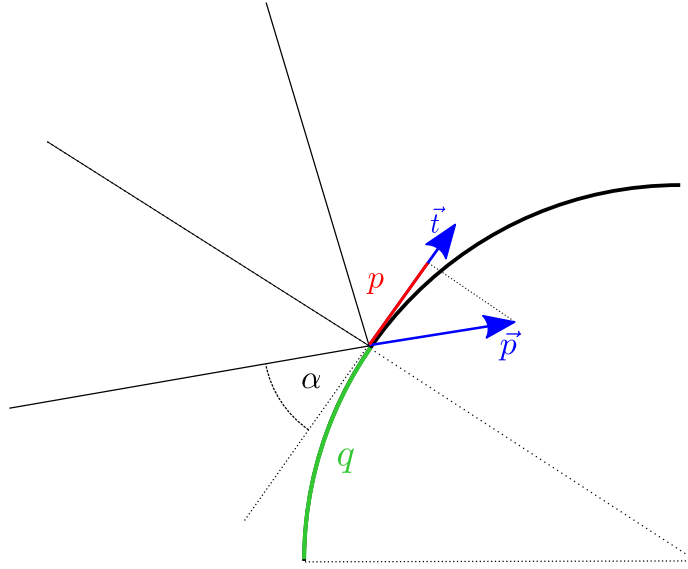}
\caption{Sketch illustrating the Birkhoff coordinates. The Birkhoff 
coordinates of a trajectory being reflected at the disk boundaries are given
by the position of the reflection on the boundary, measured by the arc length
of the impact point with respect to a fixed reference point (denoted by $q$
in the figure) and by the projection of the unit momentum vector on the unit 
tangent vector.
}\label{fig:birkhoff_coordinates}
\end{figure}

As already mentioned before the classical time evolution of a particle between two disk 
reflections is just a linear propagation interspersed with hard wall reflections. The trajectory of
a particle therefore does not depend on the absolute value of its momentum and
we can always scale it to $1$. It is furthermore convenient to get rid of the 
parts of linear propagation where no complex dynamic takes place and consider only 
the Poincar\'e section of disk reflections. A particle at 
the boundary of a disk $D_i$ of radius $a_i$ whose momentum is scaled to one, 
is completely described by its Birkhoff coordinates $(q,p)$. They are defined as
the arc length $q\in [-\pi a_i,\pi a_i]$ on the disk boundary as well as the projection
of the normalized incoming momentum to the unit tangent vector at the disk boundary 
$p=\vec p\cdot \vec t=\cos(\alpha)\in[-1,1]$ 
(see \fref{fig:birkhoff_coordinates}). The complete space
$\mathcal P$ of the Poincar\'e section consists of three copies of 
$[-\pi a_i,\pi a_i]\times[-1,1]$. As the system is open the dynamics are not defined 
on all $\mathcal P$, because in many cases a particle with Birkhoff coordinates 
$(q,p)$ will just escape to infinity after the reflection and will not hit any 
of the other disks in the future. We therefore introduce the forward 
trapped set $\mathcal T_+^{(n)}$ of order $n$ as the set of all points in 
$\mathcal P$ which will hit at least $n$ disks on their future trajectories. 
Analogously we define the backward trapped set $\mathcal T_-^{(n)}$ for the 
time inverted dynamics. The discrete time dynamics on the Poincar\'e section is 
then a diffeomorphism

\[
\phi: \mathcal T_+^{(1)}\to\mathcal T_-^{(1)}.
\]
Note that this diffeomorphism only leaves invariant the trapped set 
\[
\mathcal T = \bigcap\limits_{n=1}^\infty \mathcal T^{(n)}_+\cap\mathcal T_-^{(n)}
\]
which is known to be a fractal subset of $\mathcal P$ if the 3 disks 
are sufficiently far away from each other \cite{gas89b}. Using this discrete time 
dynamics we can define for any smooth function $V:\mathcal T_+^{(1)}\to \C$ the 
transfer operator $\mathcal L_V:C^\infty_0(\mathcal T_+^{(1)}) 
\to C^\infty(\mathcal P)$ by setting 
\[
\left(\mathcal L_Vf\right)(x) =V(\phi^{-1}(x))f(\phi^{-1}(x)) 
\]
which is an operator with integration kernel 
\begin{equation}\label{eq:kernel_transfer_op}
K(x,y)=V(y)\delta(\phi^{-1}(x)-y) .
\end{equation}
In the theory of dynamical systems this function $V$ is commonly called the
\emph{potential function} even if $V$ does not at all play the role of a potential
in the sense of classical mechanics. With this transfer operator we can now 
introduce the zeta function. If we suppose that  $\mathcal L_V$ can be extended 
to a Banach space $\mathcal B$ such that $\mathcal L_V$ is trace class
then we can study the dynamical zeta function associated to the potential $V$ which is defined by the Fredholm determinant\footnote{To our knowledge such a 
function space is not known. In physics literature such an assumption 
is, however, often implicitly made.}
\begin{equation}\label{eq:dynamical_zeta_Fredholm}
 d_{\mathrm{Fred}}(z):=\det(1-z\mathcal L_V) = \exp\left(-\sum_{n>0}\frac{z^n}{n}\Tr(\mathcal L_V^n) \right).
\end{equation}
This function is then known to be analytic in $z$ and the discrete spectrum 
$\sigma(\mathcal L_V)$  of the transfer operator is related to the zeros 
of the dynamical zeta function by
\[
 \sigma(\mathcal L_V)=\{z\setminus\{0\}\in\C,~d_{\mathrm{Fred}}(1/z)=0\}.
\]
By the explicit form (\ref{eq:kernel_transfer_op}) of the kernel $K(x,y)$ we can 
on the other hand formally define the so-called flat trace
\[
 \Tr^\flat(\mathcal L_V^n):= \sum_{x\in \mathrm{Fix}(\phi^n)}\frac{V_n(x)}{|\det(1-(D\phi^n)(x))|},
\]
where $(D\phi^n)(x)\in \R^{2\times 2}$ is the Jacobian of $\phi^n$, 
$V_n(x)=\prod_{k=1}^n V(\phi^k(x))$ the iterated product and $\tu{Fix}(\phi^n)$ 
the set of all fixed points of $\phi^n$. By the flat trace we 
can define in complete analogy to (\ref{eq:dynamical_zeta_Fredholm}) another 
zeta function by
\begin{equation}\label{eq:dynamical_zeta_flat}
d^\flat(z):= \exp\left(-\sum_{n>0}\frac{z^n}{n}\sum_{x\in \mathrm{Fix}(\phi^n)}\frac{V_n(x)}{|\det(1-(D\phi^n)(x))|} \right).
\end{equation}
By a standard calculation 
for zeta functions (see \ref{app:product_form}) this expression can be brought to the form 
\begin{equation}\label{eq:dynamical_zeta_product_form}
d^\flat(z):= \prod\limits_{p\in P} \prod\limits_{k=0}^\infty \left(1-z^{n_p} \frac{V_p}{|\Lambda_p|\Lambda_p^k}\right)^{k+1}.
\end{equation}
Here $P$ is the set of all \emph{primitive periodic orbits} of the map $\phi$, i.e.~the
set of all periodic orbits $\{x,\phi(x),\ldots \phi^m(x)=x\}$ for which there 
is no $1<k<m$ which also fulfills $\phi^k(x)=x$. Furthermore $n_p$ is the 
discrete time length of this orbit, $V_p$ is the iterated product 
of the potential $V$ along the primitive orbit $p$ and $\Lambda_p$ is 
the eigenvalue of $D\phi^{n_p}(x_p)$ whose absolute value is larger than one. 

It is important to note that the expressions (\ref{eq:dynamical_zeta_flat}) 
and (\ref{eq:dynamical_zeta_product_form}) only converge absolutely for $|z|$
sufficiently small. Furthermore (\ref{eq:dynamical_zeta_flat}) directly implies
that the zeta function has no zeros in the region of absolute convergence. 
Equations (\ref{eq:dynamical_zeta_flat}) and (\ref{eq:dynamical_zeta_product_form}) 
are thus of no direct use for a calculation of the zeros of the dynamical 
zeta function. If one knows, however, that $d^\flat(z)$ can be extended analytically 
beyond the region of absolute convergence, its Taylor expansion in $z$ around zero 
\[
 d^\flat(z)=\sum\limits_{n=0}^\infty C_n z^n
\]
will automatically converge absolutely on any disk contained in the region of analytic
continuation. This trick has been introduced in physics by Cvitanovi\'c and Eckhardt 
\cite{cvi89} under the name cycle expansion.

From a mathematical point of view, the biggest challenge is to prove the analytic
continuation of the zeta function defined by (\ref{eq:dynamical_zeta_flat}) and 
to show that its zeros correspond to the spectrum of $\mathcal L_V$. This is 
for example possible if one finds function spaces such that $\mathcal L_V$ is 
trace class and that its trace in this space equals the flat trace. Then one obtains by the 
analyticity of the Fredholm determinant automatically an analytic continuation 
of the zeta function. Such results are known for several classes of systems 
(for example for real analytic expanding or hyperbolic maps \cite{rue76,rug92, fau06})
but to our knowledge not in any setting in which the 3-disk system fits. 

Physicists usually skip this difficulty and test for concrete examples in which regions the
cycle expansion numerically converges. It is then assumed that this region coincides with
the region of analytic continuation. Furthermore it is usually assumed that the 
zeros of the zeta function defined by the flat trace (\ref{eq:dynamical_zeta_flat})
are eigenvalues of the transfer operator. As there is not yet a satisfying 
mathematical theory in the case of the 3-disk system we will also make these 
assumptions from now on.

For the 3-disk system two particular choices for the potential are of 
special interest as they lead to the classical zeta function and to the 
semiclassical or Gutzwiller-Voros zeta function, respectively. 
The return time or return length of a particle at the disk 
boundary with Birkhoff coordinates $x\in \mathcal T_+^{(1)}$ is defined as the 
length of the trajectory until the next disk reflection. This defines a smooth bounded
function
\[
\tau:\mathcal T_+^{(1)} \to \R^+ 
\] 
and we first can consider for $s\in\C$ the analytic family of potentials
\begin{equation}\label{eq:potential_classic}
 V_s^{cl}(x)=e^{-s\tau(x)}.
\end{equation}
Plugging this potential into (\ref{eq:dynamical_zeta_product_form}) we obtain 
a dynamical zeta function depending on two variables $s,z$
\[
 d^{cl}(s,z):= \prod\limits_{p\in P} \prod\limits_{k=0}^\infty \left(1-z^{n_p} \frac{e^{-s\tau_p}}{|\Lambda_p|\Lambda_p^{k}}\right)^{k+1},
\]
where $\tau_p$ is the Birkhoff sum along this orbit defined by
\[
 \tau_p=\sum_{k=1}^{n_p} \tau(\phi^k(x))
\]
for $x$ being an arbitrary point on the orbit $p$. It
is related to the Selberg-Smale zeta function by \cite{gas92}
\[
 Z(s)=d^{cl}(s,1).
\]
This zeta function is of special interest because its zeros, the so-called Ruelle resonances, 
determine the decay of correlations of the classical 
dynamical system (at least for systems where the theory of Ruelle resonances 
is established, for the 3-disk system this is again conjectured to be true, c.f.~\cite{gas92}).
This is why these zeta functions have been studied numerically by Gaspard and 
Ramirez \cite{gas92} for the symmetric 3-disk system where they observed the 
resonance chains presented in \fref{fig:res_chains}.

Using the ideas of \cite{fau13} we can also express the Gutzwiller-Voros zeta 
function by transfer operators. We therefore define the function $\Lambda(x)$
on $\mathcal T_+^{(1)}$ to be the restriction of the Jacobian $D\phi(x)$ to 
the unstable direction which is simply given by the multiplication with a 
nonzero number. If $x$ is a fixed point of an iteration of the Poincar\'e map
$\phi^n(x)=x$, then the iterated product of $\Lambda(x)$ will be exactly the 
unstable eigenvalue $\Lambda_n$ of $D\phi^n$. 

With this function we define
\[
 V_s^{(a)}(x) = -|\Lambda(x)|^{1/2}e^{-s\tau(x)}\tu{ and }V_s^{(b)}(x) = |\Lambda(x)|^{-1/2}e^{-s\tau(x)}
\]
as well as the corresponding Fredholm determinants 
\[
d^{(a)}(s,z) =\det(1-z\mathcal L_{V_s^{(a)}}) \tu{ and } d^{(b)}(s,z) =\det(1-z\mathcal L_{V_s^{(b)}}).
\]
We can then study the quotient
\begin{eqnarray}\label{eq:Gutzwiller_Voros_zeta}
 \frac{d^{(a)}(s,z)}{d^{(b)}(s,z)}&= \exp\left( -\sum_{n>0}\frac{z^n}{n}\sum_{x\in \mathrm{Fix}(\phi^n)}\frac{e^{-s\tau_n(x)-in\pi}\left(|\Lambda_n(x)|^{1/2}-(-1)^n|\Lambda_n(x)|^{-1/2}\right)}{|\det(1-(D\phi^n)(x))|} \right) \nonumber\\
 &=\exp\left( -\sum_{n>0}\frac{z^n}{n}\sum_{x\in \mathrm{Fix}(\phi^n)}\frac{e^{-s\tau_n(x)-in\pi}}{\sqrt{|\det(1-(D\phi^n)(x))|}}\right).\nonumber
\end{eqnarray}
For the last equality we used that the Poincar\'e map is orientation inverting 
and consequently $\Lambda(x) <0$ which implies that  
\[
\sqrt{|\det(1-(D\phi^n)(x))|}=|\Lambda_n(x)|^{1/2}-(-1)^n|\Lambda_n(x)|^{-1/2}. 
\]
The final expression is related to the so-called Gutzwiller-Voros zeta function \cite{vor88}
by
\begin{equation}\label{eq:GV_Transfer_equivalence}
 Z_{GV}(k)=\frac{d^{(a)}(-ik,1)}{d^{(b)}(-ik,1)}.
\end{equation}
It has been 
numerically thoroughly checked \cite{cvi89,wir99b} that the zeros of $Z_{GV}(k)$
coincide to a high 
precision to the quantum resonance which have numerically been obtained by a 
quantum scattering approach. It is conjectured that the zeros of the 
Gutzwiller-Voros zeta function converge towards the quantum resonances in the 
semiclassical limit. This relation \cite{gas89a} is, however, based on the 
Gutzwiller-trace formula 
for the quantum propagator beyond the Ehrenfest time, and from the mathematical 
point of view the relation to the quantum spectrum is completely open
\footnote{For closed systems there has recently been derived a relation between 
the quantum and the so-called prequantum operator which is an important step into this direction \cite{arXfau12}.}.
For practical purpose this relation between
the zeros of the zeta function and the quantum resonances has, however, a huge 
advantage because especially in the regime of high energies the calculations 
of the zeros of $Z_{GV}(k)$ via the cycle expansion is much faster than the
calculation based on the quantum scattering matrix and it has been exploited in 
many numerical works on the quantum spectrum \cite{lu03, ebe10, wie08}.

\subsection{Rotating resonances}\label{sec:rot_res}
We now want to understand the resonance chains for the symmetric 3-disk 
system via the zeta functions introduced above. The zeta functions allow to 
handle the resonance chains for classical Ruelle resonances and quantum 
resonances by the same approach, even though the physical nature of these resonances
is completely different. 

In both cases one has a zeta function $Z(s,z)$ which depends on two complex variables
and the resonances are defined by 
\footnote{In the case of quantum resonances 
the resonance set is usually the set $\tu{Res}=\{k\in \C,~Z(-ik,1)=0\}$ 
(see (\ref{eq:GV_Transfer_equivalence})). In order to emphasize the uniform
approach to quantum and classical resonances as well as the geometric resonances
on Schottky surfaces, we will, however, from now on consider the quantum resonances
to be parametrized by $s=-ik$ which simply leads to a clockwise rotation by $90^{\circ}$ of the quantum resonances in the plots. 
} 
\[
\tu{Res}=\{s\in \C,~Z(s,1)=0\}.
\]
Note that for the quantum case we use the hypothesis that the semiclassical and 
quantum resonances coincide. We will, however, only use this correspondence for 
the symmetric 3-disk system in an energy regime, where this correspondence is 
known to hold with a very high precision \cite{cvi89,wir99b}. Beside
the set of resonances which we have defined to be the zeros in $s$ of $Z(s,z)$
for fixed $z=1$ it seems also natural to consider for a fixed $s\in \C$ the 
zero set in $z$. In order to have a direct physical interpretation of this set
we consider the inverses of these zeros and define for a fixed $s\in \C$
\begin{equation}\label{eq:sigma_s_3-disk}
 \sigma_s:=\left\{z\in\C\setminus\{0\},~Z(s,1/z)=0\right\}.
\end{equation}
Using the Fredholm determinant representation of the classical zeta function
\[
 Z_{cl}(s,z)=\det(1-z\mathcal L_{V_s^{cl}})
\]
one directly identifies $\sigma_s=\sigma (\mathcal L_{V_s^{cl}})$ which is
the spectrum of $\mathcal L_{V_s^{cl}}$. For the semiclassical resonances this
identification is slightly more subtle as $Z_{GV}$ has been defined as the 
quotient of two zeta functions. If we assume, however, that $d^{(b)}(s,z)$
is analytic in a neighborhood of $z\in \sigma_s$, then the Gutzwiller-Voros
zeta function can only vanish if the nominator of the quotient vanishes, 
i.e.~if
\[
 0=d^{(a)}(s,z^{-1})=\det(1-z^{-1}\mathcal L_{V_s^{(a)}})
\]
and $z\in \sigma(\mathcal L_{V^{(a)}_s})$. Note that the analyticity assumption
of $d^{(b)}(s,z)$ is not very strong and in the $s$-range which we will 
consider in the sequel it is always fulfilled (see \ref{app:top_pres})

Summarizing, we have the following two objects in the classical as well as in 
the semiclassical or quantum case: On the one hand we have the resonance
spectrum which equals the zeros of the zeta function in $s$ for $z=1$ and on the
other hand we have for any $s\in \C$ the discrete set $\sigma_s$ which can be 
interpreted as the eigenvalues of a transfer operator $\mathcal L_s$ which 
is parametrized by a complex number $s$.
Obviously $s$ is a resonance if and only if 
the operator $\mathcal L_{s}$ has $1$ as an eigenvalue i.e.
\[
 s\in \tu{Res} \Leftrightarrow 1\in \sigma_{s}.
\]
The simple but important observation is now the following: If we start with a
resonance $s_0\in \tu{Res}$ of multiplicity one and continuously vary the 
parameter $s$ then we will be able to trace the eigenvalue which was equal to $1$ for 
$s=s_0$ and which continuously
depends on $s$. We will call this eigenvalue $\lambda_{s_0}(s)$.
A closer look in the concrete form of our transfer operators 
$\mathcal L_s$ allows us to predict how the eigenvalues will approximately 
depend on $s$. In the quantum case for $\mathcal L_s=\mathcal L_{V^{(a)}_s}$ as
well as in the classical case with $\mathcal L_s=\mathcal L_{V_s^{cl}}$ we have
\[
 \mathcal L_s= \mathcal L_0e^{-s\tau(x)}
\]
where $e^{-s\tau(x)}$ is simply the multiplication operator. In the case of
the symmetric 3-disk system the return time is a smooth function
which becomes more and more homogeneous the bigger $R/a$ becomes.
Assuming as a first order approximation that $\tau(x)\approx \tau_0=R-2a$ then
\[
 \lambda_{s_0}(s) = e^{-s\tau_0}.
\]
If this approximation was exactly true and if we moved with $s$ parallel to the 
imaginary axis, then the eigenvalue $\lambda_{s_0}(s)$ would simply rotate around zero 
on the unit circle. Note that by the definition of the resonance set,  
every $s$ value for which $\lambda_{s_0}(s)$ crosses the value point 1 on the unit circle
is itself a resonance. By this argument a single eigenvalue of $\mathcal L_{s_0}$
would create a whole chain of resonances, parallel to the imaginary axis.The 
resonances would be equally distributed along this chain with a distance of $2\pi/\tau_0$.
As the return time is in reality not exactly constant one would expect that 
slight variations deform the chains and also lead to a distance variation between 
the extremal values of the return time. 

\begin{figure}
\centering
        \includegraphics[width=\textwidth]{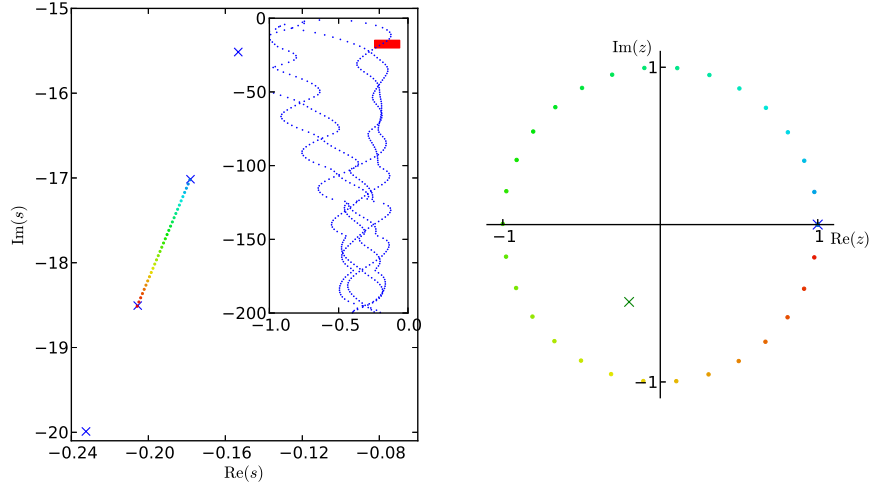}
\caption{Left: Plot of the quantum resonances of the 3-disk system 
with $R/a=6$. The inset shows 
the chain structure of the resonances on a large domain in the complex 
plane. The main plot shows a strong zoom into one chain, such that
only few individual resonances are in the plot region (blue crosses). The
region of the zoom is indicated in the inset by a red rectangle. The 
colored dots between the two resonances $s_1=-0.206-18.51i$ and 
$s_2=-0.178-17.02i$ represent a discrete interpolation between the two 
resonances. The interpolation follows the line suggested by the resonance chains. Right:
Spectrum of the transfer operator on the Poincar\'e section. The blue 
crosses represent the spectrum of $\mathcal L_{s_1}$ with $|z|>0.3$. 
By the colored dots we follow the evolution of the eigenvalue 
starting at 1 for $s_1$ while $s$ varies as indicated by the 
colored points in the left panel. The evolution of the second eigenvalue 
is not plotted for more clarity.}
\label{fig:rot_res_semiclassic}
\end{figure}

\begin{figure}
\centering
        \includegraphics[width=\textwidth]{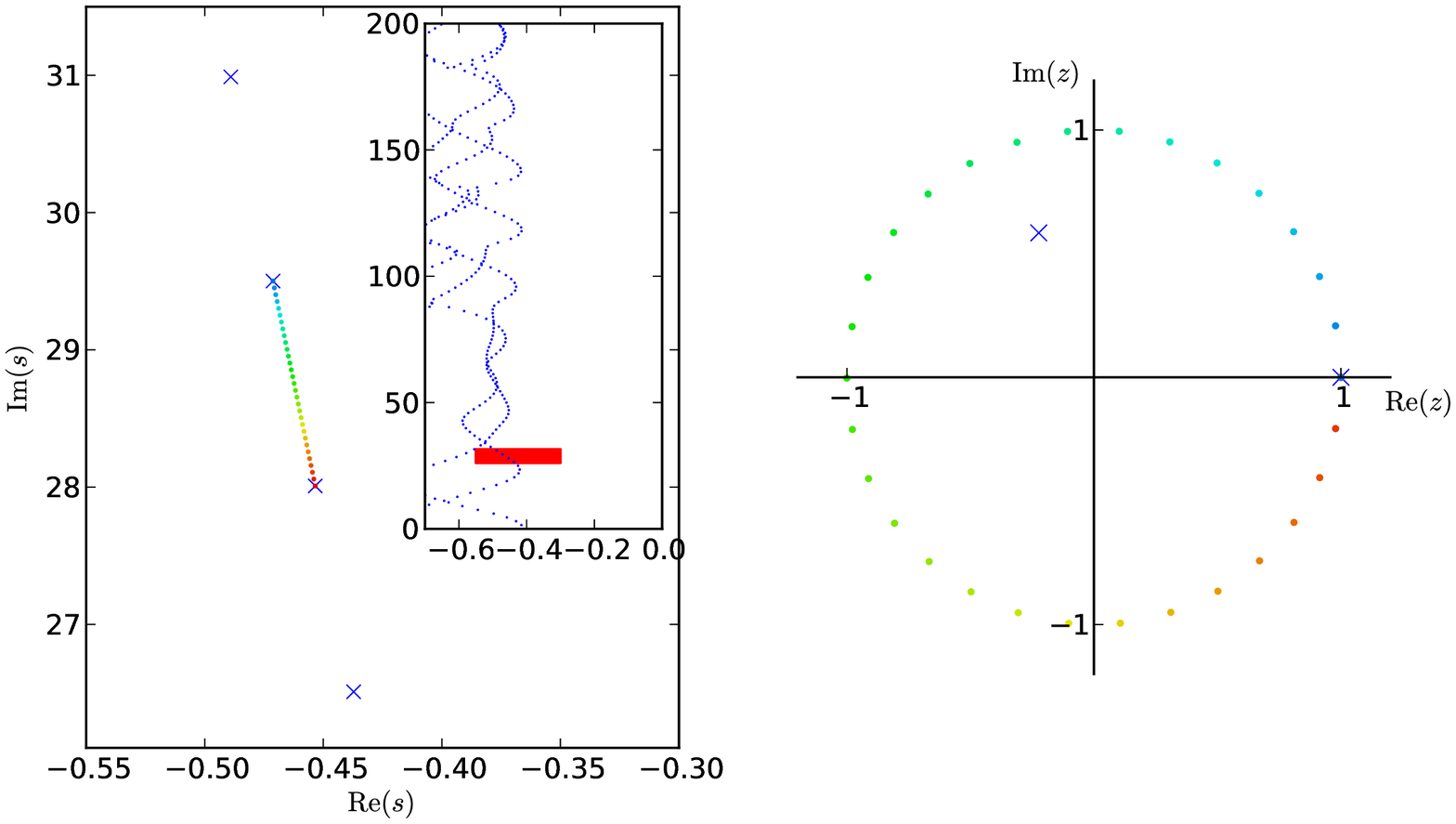}
\caption{Left: Plot of the classical Ruelle resonances of the 3-disk system 
with $R/a=6$. The inset shows 
the chain structure of the resonances on a large domain in the complex 
plane. The main plot shows a strong zoom into one chain, such that 
only few individual resonances are in the plot region (blue crosses). The
region of the zoom is indicated in the inset by a red rectangle. The 
colored dots between the two resonances $s_1=0.453+28.0i$ and 
$s_2=0.471+29.5i$ represent a discrete interpolation between the two 
resonances following the line suggested by the resonance chains. Right:
Spectrum of the transfer operator on the Poincar\'e section. The blue 
crosses represent the spectrum of $\mathcal L_{s_1}$ with $|z|>0.3$. 
By the colored dots we follow the evolution of the eigenvalue 
starting at 1 for $s_1$.}
\label{fig:rot_res_classic}
\end{figure}

It can be checked numerically that it is exactly this mechanism of a 
rotating eigenvalue of the operator $\mathcal L_s$ which creates the resonance chains.
In \fref{fig:rot_res_semiclassic} this is visualized for the quantum resonances
of the symmetric 3-disk system with $R/a=6$. On the left we can see a plot of the
quantum resonances $s_n$ in the complex plane\footnote{Note that as discussed above the
usual parametrization of quantum resonances is $k_n=is_n$.}. On the right
the spectrum of the transfer operator $\mathcal L_{V_s^{(a)}}$ is presented. The blue crosses indicate
the eigenvalues with $|z|>0.3$ for an $s$-value which equals the resonance $s_1=-0.206-18.51i$.
As expected we find an eigenvalue which is equal to $1$. There is also another
eigenvalue plotted which corresponds to the second chain which is nearby the 
chain to which $s_1$ belongs. There are further eigenvalues of the transfer 
operator near zero which we did not plot as the convergence of the zeta function 
becomes more and more complicated the smaller the absolute values of $z$. The colored 
points show how the eigenvalue at $1$ moves, if the $s$ values varies between
two resonances $s_1$ and $s_2=-0.178-17.02i$. As suggested by the heuristic 
arguments above the eigenvalue moves 
along the unit circle in a clockwise orientation and creates the next 
resonances on the chain. Exactly the 
same can be observed for the classical Ruelle resonances of the 3-disk system 
with $R/a=6$ (see \fref{fig:rot_res_classic}).

\begin{figure}
\centering
        \includegraphics[width=\textwidth]{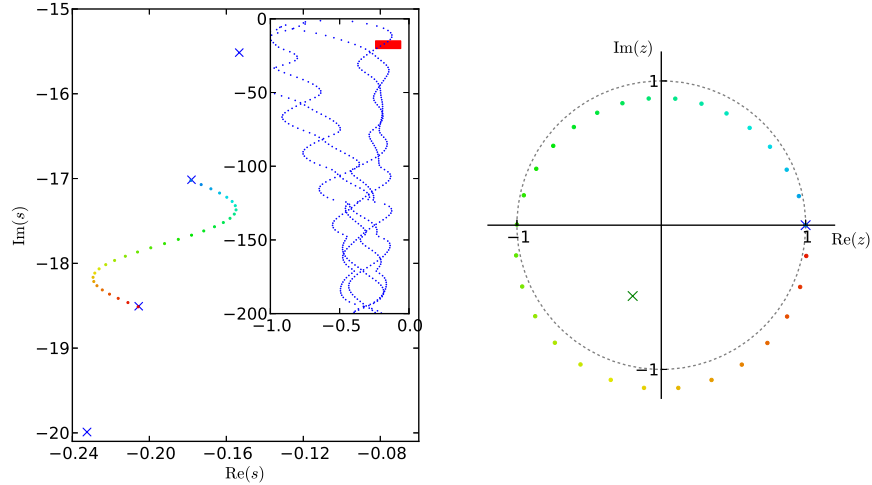}
\caption{Same as in \fref{fig:rot_res_semiclassic}. However the 
colored dots between the two resonances $s_1=-0.206-18.51i$ and 
$s_2=-0.178-17.02i$ now do not follow the line which is suggested
by the resonance chains but a deformed path. As a consequence 
the eigenvalue of $\mathcal L_{s_1}$ starting at 1 for $s_1$ now turns
one time around zero on a path which is not on the unit circle (black,dotted).}
\label{fig:rot_res_semiclassic_sinus}
\end{figure}

These numerical results do, however, not only confirm that the hypothesis that 
each resonance chain is created by a rotating $\mathcal L_s$-eigenvalue is true. They
also indicate what the continuous mathematical object is that the resonances
are distributed on. Note that taking two neighboring resonances of one chain  
there are multiple paths to connect these two resonances. While each path will 
effectively lead to one rotation of $\lambda_{s_0}(s)$ around zero, there is only
one such path, for which $\lambda_{s_0}(s)$ stays on the unit circle. As it can 
be seen in \fref{fig:rot_res_semiclassic} such a rotation on the unit circle is
obtained if the two resonances are connected along the path which is suggested by
the structure of the resonance chains. If one, however, takes a different path 
between the two resonances, then the corresponding eigenvalue of $\mathcal L_s$
will turn on a deformed circle around zero 
(see \fref{fig:rot_res_semiclassic_sinus} for an arbitrary path between the two resonances). We can thus define
\[
 \mathcal C:=\{(s,z)\in \C^2, Z(s,z)=0, |z|=1\}\subset \C^2
\]
and from the numerics above we conclude that the continuous object on which the 
resonances are distributed is given by $\tu{Pr}_s(\mathcal C)$ where 
$\tu{Pr}_s:\C^2\to\C$ is the projection on the $s$-component. 

We have thus seen in this section that the resonance chains of the classical and 
the quantum resonances are generated by the same mechanism. As in both cases the 
resonances can be seen as the zero set in $s$ of a zeta function, which can be represented
as a Fredholm determinant of a $s$-dependent family of operators. The resonance
chains are generated by single eigenvalues of $\mathcal L_s$ which circle
around zero while $s$ moves along the chain. Furthermore this approach allows 
a system-intrinsic definition of a continuous object that coincides with the 
visually suggested curves on which the resonances are strung together. 
In the next section we will see how this idea can be extended to the resonance
chains on Schottky surfaces and examine in detail the question in which systems
resonance chains will occur and in which not.

\section{Schottky surfaces and generalized zeta functions}
\subsection{Introduction to Schottky surfaces}\label{sec:intro_Schottky}
Schottky surfaces are a special class of infinite volume Riemann surfaces. 
In this section we will give a brief introduction to its most important
properties. For a detailed introduction we refer to \cite{bor07}. 

The easiest way to define Schottky surfaces is as a quotient of the upper half plane by
a so-called Schottky group. The upper half plane 
$\mathbb H =\{u=x+iy\in \C,~y>0\}$ together with 
the Riemannian metric $(dx^2+dy^2)/y^2$ has a constant negative Gauss curvature 
$\kappa =-1$ and is a standard model of hyperbolic geometry. The group $SL(2,\R)$ of 
real invertible matrices with determinant one acts on $\mathbb H$
via Moebius transformations
\begin{equation}\label{eq:moebis}
 \left(\begin{array}{cc} a&b\\c&d\end{array}\right)u:=\frac{au+b}{cu+d}.
\end{equation}
In fact for every $A\in SL(2,\R)$ the Moebius transformation is an orientation
preserving isometry and all orientation preserving isometries of the upper 
half-plane can be expressed as an $SL(2,\R)$-action. A Schottky group is then
a discrete subgroup $\Gamma\subset SL(2,\R)$ which is constructed in the 
following way: Let $D_1,\ldots,D_{2r}$ be disks with centers on the real line
and mutually disjoint closure. Then for every $1\leq i\leq r$ there exists an
element $S_i\in SL(2,\R)$ that maps the boundary $\partial D_i$ to the boundary
of $\partial D_{i+r}$ and the interior of $D_i$ to the exterior of $D_{i+r}$. 
From the disjointness of the disks follows that the elements $S_1,\ldots,S_r$
are the generators of a free discrete subgroup
\[
\Gamma=\langle S_1,\ldots,S_r\rangle\subset SL(2,\R) 
\]
and such groups are called \emph{Schottky groups}. The quotient
\[
 X=\Gamma\backslash\mathbb H
\]
is then again a surface with constant negative curvature and is called
\emph{Schottky surface}.

\begin{figure}
\centering
        \includegraphics[width=0.7\textwidth]{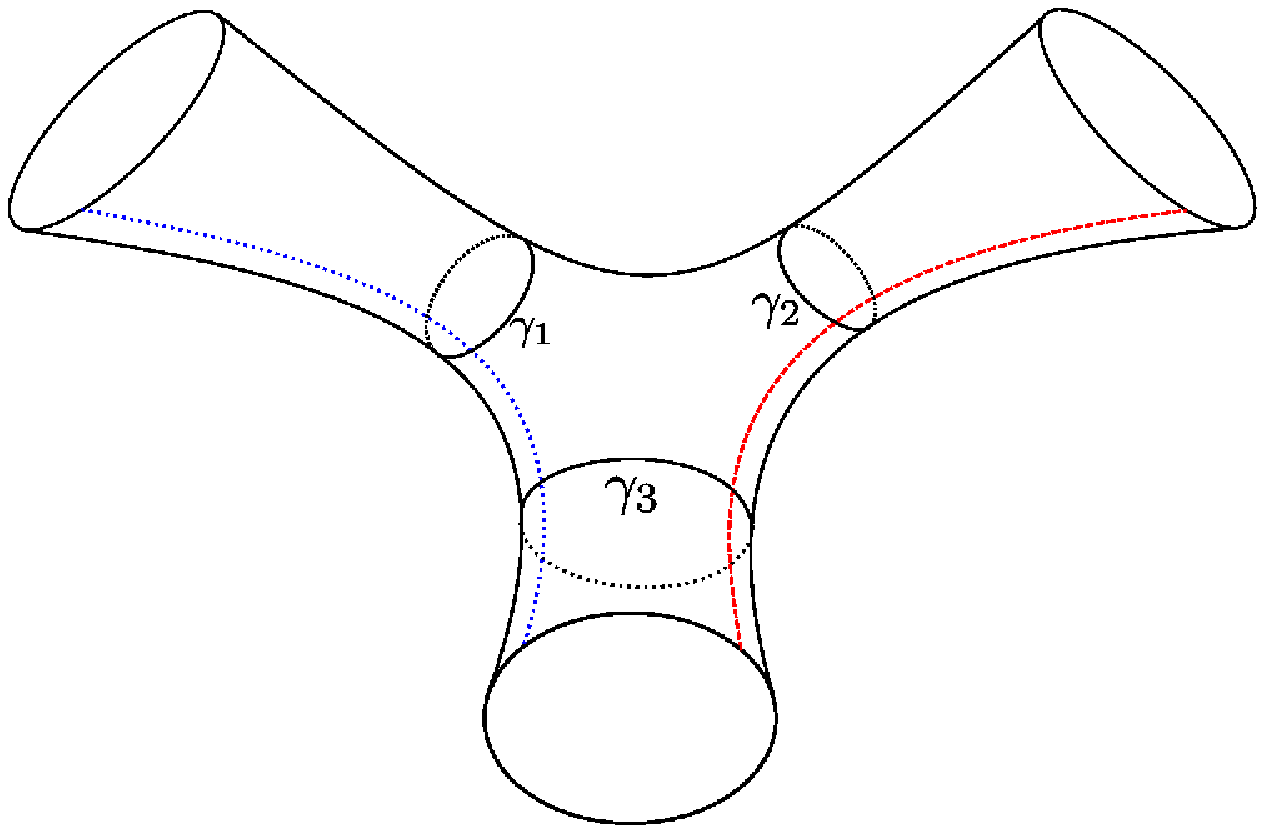}
        \includegraphics[width=\textwidth]{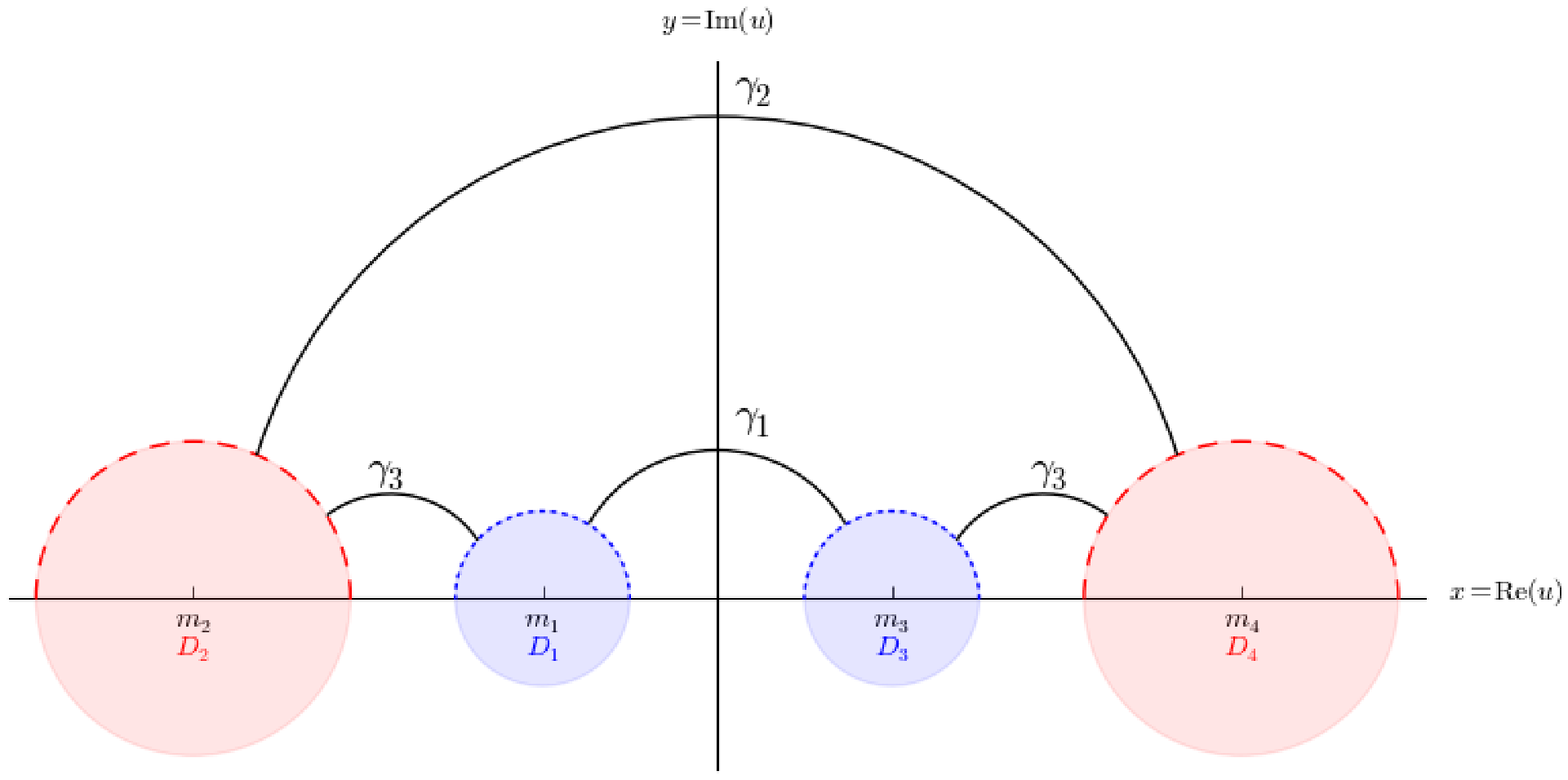}
\caption{Upper part: Sketch of a Schottky surface with three funnels and the three
fundamental closed geodesics $\gamma_1,\gamma_2,\gamma_3$. The dotted blue and 
dashed red lines indicate the lines along which the fundamental domain is glued 
together in order to obtain the surfaces. Lower part: Configuration of 4 disks that
give rise to the construction of the Schottky group for a 3-funneled surface. 
The upper half plane
without the disks represents a fundamental domain and the surface can be obtained by gluing
together the red dashed lines with the blue dotted lines. In black the three fundamental
closed geodesics $\gamma_1,\gamma_2,\gamma_3$ from the upper part of the figure are shown. 
While $\gamma_1$ and $\gamma_2$ are only represented by one arc each, the geodesic $\gamma_3$
appears as two arcs in the fundamental domain.}
\label{fig:schottky}
\end{figure}

The 3-funneled surfaces, which we are interested in, are from the 
dynamical point of view the simplest nontrivial example and they are constructed
from four disks arranged as in \fref{fig:schottky}. The surface can be thought as the set
$\mathbb H\setminus \cup D_i$ glued together along the two dotted blue half circles
as well as along the two dashed red half circles (see \fref{fig:schottky}). It
is known that such 3-funneled Schottky surfaces are uniquely determined
by the three lengths $l_1,l_2,l_3$ of the geodesics $\gamma_1,\gamma_2,
\gamma_3$ that wind once around one of the 3-funnels (see upper part of
\fref{fig:schottky}). Furthermore for any triple of positive numbers 
$l_1,l_2,l_3$ there exists a 3-funneled Schottky surface which 
has fundamental geodesics of this length and we will call this surface 
$X_{l_1,l_2,l_3}$.

In order to define the geometric resonances on a Schottky surface $X$ we consider
the positive Laplacian which is defined by the metric and which we 
call $\Delta_X$ and we define the resolvent 
\[
 R_X(s) := (\Delta_X-s(s-1))^{-1}.
\]
By Mazzeo-Melrose \cite{maz87} and Guillope-Zworski \cite{gui97} we know that the resolvent 
admits a meromorphic continuation and we define the resonance set $\tu{Res}_X$ 
again as the set of its poles. Note that the resolvent's parametrization is 
different from the parametrization via $k$ in the 3-disk case, which leads 
to the fact that the resonance chains do not oscillate along the real axis
but along  the imaginary axis. This parametrization is, however, very useful as it 
admits the following exact relation between the resonances
of the Laplacian and a zeta function:

Let $P_X$ be the set of all primitive closed geodesics on $X$, i.e.~those
closed geodesics that cannot be obtained by repeting a shorter closed
geodesic. If
for $\gamma\in P_X$, $l(\gamma)$ denotes the length of the periodic geodesic, then the
Selberg zeta function is defined as 
\begin{equation}\label{eq:sel_zeta}
 Z_X(s) = \prod_{\gamma\in P_X}\prod_{m=0}^\infty \left(1-e^{-(s+m)l(\gamma)}\right).
\end{equation}
This product is known to converge uniformly on any compact set with $\tu{Re}(s)>1$ and for
$X$ being a Schottky surface the Selberg zeta function is known to extend 
analytically to the whole complex plane $\C$ and there is the following relation
between its zeros and the resonances of $\Delta_X$: If $Z_X(s)=0$ then $s$ is 
either a topological zero with $s=0,-1,-2,\ldots$ or a resonance 
i.e.~$s\in \tu{Res}_X$ \cite{pat01}.

For Schottky surfaces we can express the Selberg zeta function as a Fredholm
determinant of a transfer operator. We therefore define the 
\emph{Bowen-Series map} of a Schottky group $\Gamma=\langle S_1,\ldots,S_r\rangle$ by 
\[
 B:\Abb{\bigcup\limits_{i=1}^{2r} D_i}{\C}{u}{S_iu \tu{ if }u\in D_i.}
\]
Here we used the convention that for $r<i\leq 2r$, $S_i:=S_{i-r}^{-1}$. We can 
then define a family of transfer operators parametrized by $s\in \C$ via its action
\[
 \left(\mathcal L_s f\right) (u):= \sum\limits_{v\in B^{-1}(u)} B'(v)^{-s}f(v),
\]
where $B'$ is the complex derivative of the Bowen-Series map and $u\in \cup_i D_i$.
These transfer operators are known to be trace class on the space of holomorphic
$L^2$-functions on $\cup_i D_i$ (see \cite{rue76} for the original proof in slightly 
different functions spaces or \cite[Lemma 15.6]{bor07}) and one therefore knows that its dynamical 
zeta function
\[
 d_{BS}(s,z):=\det\left(1-z\mathcal L_s\right)
\]
is analytic. We will give a short sketch of the idea how to relate the dynamical 
zeta function to the Selberg zeta function. For more details we refer to 
\cite[Section 15.3]{bor07}. One first can proof that the trace of $\mathcal L_s$ 
equals its flat trace and obtains
\begin{equation}
 \label{eq:BS_dyn_zeta_fixpoint_formula}
 d_{BS}(s,z)=\exp\left(-\sum_{n>0}\frac{z^n}{n}\sum_{u\in \mathrm{Fix}(B^n)}\frac{(B^n)'(u)^{-s}}{|1-(B^{-n})'(u)|} \right).
\end{equation}
As in the case of the 3-disk system (cf.~equation (\ref{eq:dynamical_zeta_product_form}) and \ref{app:product_form})
one can transform the dynamical zeta function into a product over prime orbits of the 
Bowen-Series map $B$. In a second step one can prove that the set of primitive 
periodic geodesics is in one-to-one correspondence to the 
primitive periodic orbits of the Bowen-Series maps, i.e.~for each $\gamma\in P_X$
there exists a unique $n\in \N$ and a periodic orbit $\{u,B(u),\ldots,B^n(u)=u\}$ 
and the length of the geodesic is related to the stability of the fixed point
by
\[
e^{l(\gamma)}=(B^n)'(u).
\]
If we denote this $n$ as the \emph{Bowen-Series order} $n_{BS}(\gamma)$ of the 
geodesic $\gamma$, then we obtain, combining all arguments
\begin{equation}\label{eq:dyn_zeta_BS}
 d_{BS}(s,z) = \prod_{\gamma\in P_X}\prod_{m=0}^\infty \left(1-z^{n_{BS}(\gamma)}e^{-(s+m)l(\gamma)}\right).
\end{equation}
In particular this yields $d_{BS}(s,1)=Z_X(s)$ and we deduce that 
$s\in\tu{Res}_X$ implies that $\mathcal L_s$ has an eigenvalue equal to 1.

\subsection{Generalized zeta function and spectra}\label{sec:gen_zeta}
\begin{figure}
\centering
        \includegraphics[width=\textwidth]{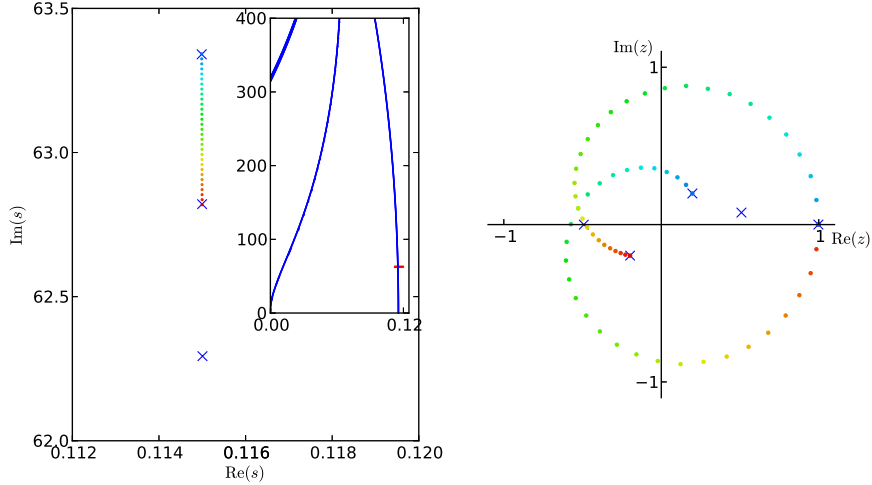}
\caption{Left: Plot of the resonances of $X_{12,12,12}$. The inset shows 
the chain structure of the resonances on a large domain in the complex 
plane. The main plot shows a strong zoom into one chain, such that the 
individual resonances become visible (blue crosses). The
region of the zoom is indicated in the inset by a red rectangle. The 
colored dots between the two resonances $s_1=0.1150+62.82i$ and 
$s_2=0.1150+63.34i$ represent a discrete interpolation between the two 
resonances, following the line suggested by the resonance chains. Right:
Spectrum of the Bowen-Series transfer operator with $|z|>0.2$. 
The blue crosses represent 
the spectrum of $\mathcal L_{s_1}$. By the colored dots we follow the 
evolution of two particular eigenvalues, as the $s$-values evolve as
indicated by the colored points in the left panel. We only follow the 
evolution of the eigenvalue starting at 1 for $s_1$ and the one ending at 1 
for $s_2$. The evolution of the other eigenvalues is not shown for more
clarity.}
\label{fig:rot_res_BS}
\end{figure}

The previous discussion on the relation between the spectrum of $\mathcal L_s$ 
and the resonances of the Laplacian, together with the observations from Section~\ref{sec:rot_res}
that for 3-disk systems rotating eigenvalues of a transfer operator create the resonance
chains would suggest that the 
resonance chains for the 3-funneled Schottky surfaces are generated
by rotating eigenvalues of the Bowen-Series transfer operator 
$\mathcal L_s$. However, this idea fails completely to explain the 
resonance chains for the 3-funneled symmetric surfaces: 
In \fref{fig:rot_res_BS} we have plotted for $X_{12,12,12}$ the spectrum 
of $\mathcal L_s$ for different $s$-values interpolating between two 
resonances $s_1$ and $s_2$.
 We observe that the eigenvalues start to turn 
while we move along the chain. However, the eigenvalue which was equal to 
$1$ for $s=s_1$ is drawn towards zero and when we approach $s=s_2$ it is 
another eigenvalue of $\mathcal L_s$ which has been coming out from the 
interior, that creates the next resonance. In order to understand why a 
direct application of the idea of rotating eigenvalues to the spectrum of 
the Bowen-Series transfer operator $\mathcal L_s$ fails, we recall
that for the 3-disk system it has been an important argument that
the return time of the Poincar\'e section is very homogeneous. This
assumption is, however, not at all fulfilled for the standard Bowen-Series map,
which corresponds morally to a Poincar\'e section\footnote{To be more precise, a 
Poincar\'e section along these two lines would by a bijective hyperbolic
map on a two dimensional space. The Bowen-Series map can then be obtained
as the restriction of this Poincar\'e map to the instable direction, which 
makes it expanding but not invertible anymore.} indicated by the blue
and red lines in \fref{fig:schottky}. The return time or return length respectively
is thus very inhomogeneous: While on $X_{12,12,12}$ for the closed geodesics $\gamma_1$ and 
$\gamma_2$ the return length is equal to $12$, it equals only $6$ for the geodesic 
$\gamma_3$ as this geodesic crosses the Poincar\'e section twice. 

\begin{figure}
\centering
        \includegraphics[width=\textwidth]{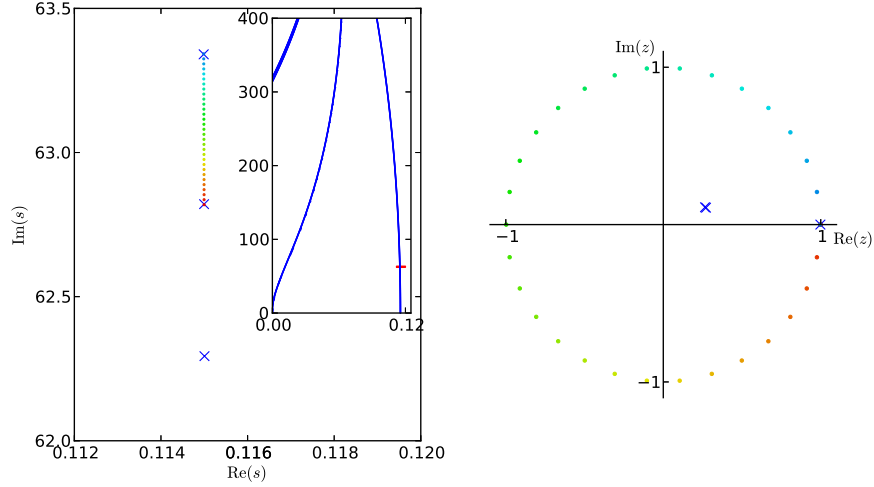}
\caption{Left: Same as in \fref{fig:rot_res_BS}. Right:
Plot of the generalized spectrum $\sigma_s^{(\mathbf n)}$ with $|z|>0.2$
for an order function
that counts the number of windings around the funnels. The blue 
crosses represent the spectrum $\sigma_{s_1}^{(\mathbf n)}$. By the colored dots 
we follow the evolution of the spectral value which equals 1 for 
$s=s_1$. The evolution of the other spectral values is again not plotted.}
\label{fig:rot_res_GZ}
\end{figure}

In order to fix this issue we could try to construct another transfer operator
which still has the property that its dynamical zeta function is related to the 
Selberg Zeta function but which has a homogeneous return time for the completely
symmetric 3-funneled Schottky surface. Such a construction is in principle
possible (see \cite{mathArticle}), this solution would then, however, be restricted to completely
symmetric Schottky surfaces. Non-symmetric surfaces, for which resonance chains have
also been observed, would then need a construction of more and more complicated 
transfer operators. We will thus choose a different approach and work directly
with the zeta functions which turns out to be much more flexible. Looking at equation
(\ref{eq:dyn_zeta_BS}) the choice of the Poincar\'e section appears only in the order
function $n_{BS}$. This order function has been defined as the length of the prime
$B$-orbit which is associated to the primitive geodesic $\gamma$ and it is exactly 
given by the number of times the geodesic passes one of the cut lines in \fref{fig:schottky}.
Instead of constructing another Poincar\'e section we can also directly modify the zeta 
function and we can introduce for each map
\[
 \mathbf n: P_X \to\N
\]
the \emph{generalized zeta function} 
\begin{equation}\label{eq:gen_zeta}
 d_{\mathbf n}(s,z):= \prod_{\gamma\in P_X}\prod_{m=0}^\infty \left(1-z^{\mathbf n(\gamma)}e^{-(s+m)l(\gamma)}\right).
\end{equation}
Of course for a general order function $\mathbf n$ it is not directly clear 
where this product converges and whether it has an analytic extension.
For all order functions that will appear in the sequel, one can, however, show, 
that such an analytic extension exists and we refer to \cite{mathArticle} for a rigorous
proof. The important observation is now that changing the order function only 
changes the values of the dynamical zeta functions for $z\neq 1$ and 
the important relation 
\begin{equation}\label{eq:gen_zeta-Selberg_zeta}
d_{\mathbf n}(s,1)=Z_X(s)
\end{equation}
is still true. 
We can therefore define in analogy to the spectrum of the transfer operator (\ref{eq:sigma_s_3-disk}), 
the \emph{generalized spectrum}
\begin{equation}\label{eq:gen_spectrum}
 \sigma_s^{(\mathbf n)}:=\left\{z\in\C\setminus\{0\},~d_{\mathbf n}\left(s,1/z\right)=0 \right\}.
\end{equation}
From (\ref{eq:gen_zeta-Selberg_zeta}) we conclude that also for this 
generalized spectrum we have the important relation
\[
 s\in \tu{Res}_X \Rightarrow 1\in \sigma_s^{(\mathbf n)}.
\]
Thus instead of tracing the eigenvalues of a transfer operator we
can follow the generalized spectral values while interpolating between
different resonances on one chain. The order function which would
correspond to a Poincar\'e section of the symmetric surface $X_{12,12,12}$
with a homogeneous return time counts for each closed geodesic how often it winds around one of 
the funnels. In \fref{fig:rot_res_GZ} we traced the generalized spectrum
while we interpolate between the same two resonances as in 
\fref{fig:rot_res_BS} and indeed we observe that the generalized 
spectral value which equals $1$ for $s=s_1$ starts to turn around zero and
creates all the other resonances on this chains by passing through $z=1$.
Additionally we see in \fref{fig:rot_res_GZ} that if the $s$-values follow
the path which is suggested by the resonance chains, then the generalized spectral 
value stays on the unit circle. The continuous lines on which the resonances lie
can thus be understood by the real analytic variety
\begin{equation}
 \label{eq:generalized_chain_variety}
 \mathcal C_{\mathbf n} = \{(s,z)\in\C^2, d_{\mathbf n}(s,z)=0,|z|=1\}.
\end{equation}
Precisely they are again given by $\tu{Pr}_s(\mathcal C_{\mathbf n})$, the projection of this variety on the 
$s$-component.

\begin{figure}
\centering
        \includegraphics[width=\textwidth]{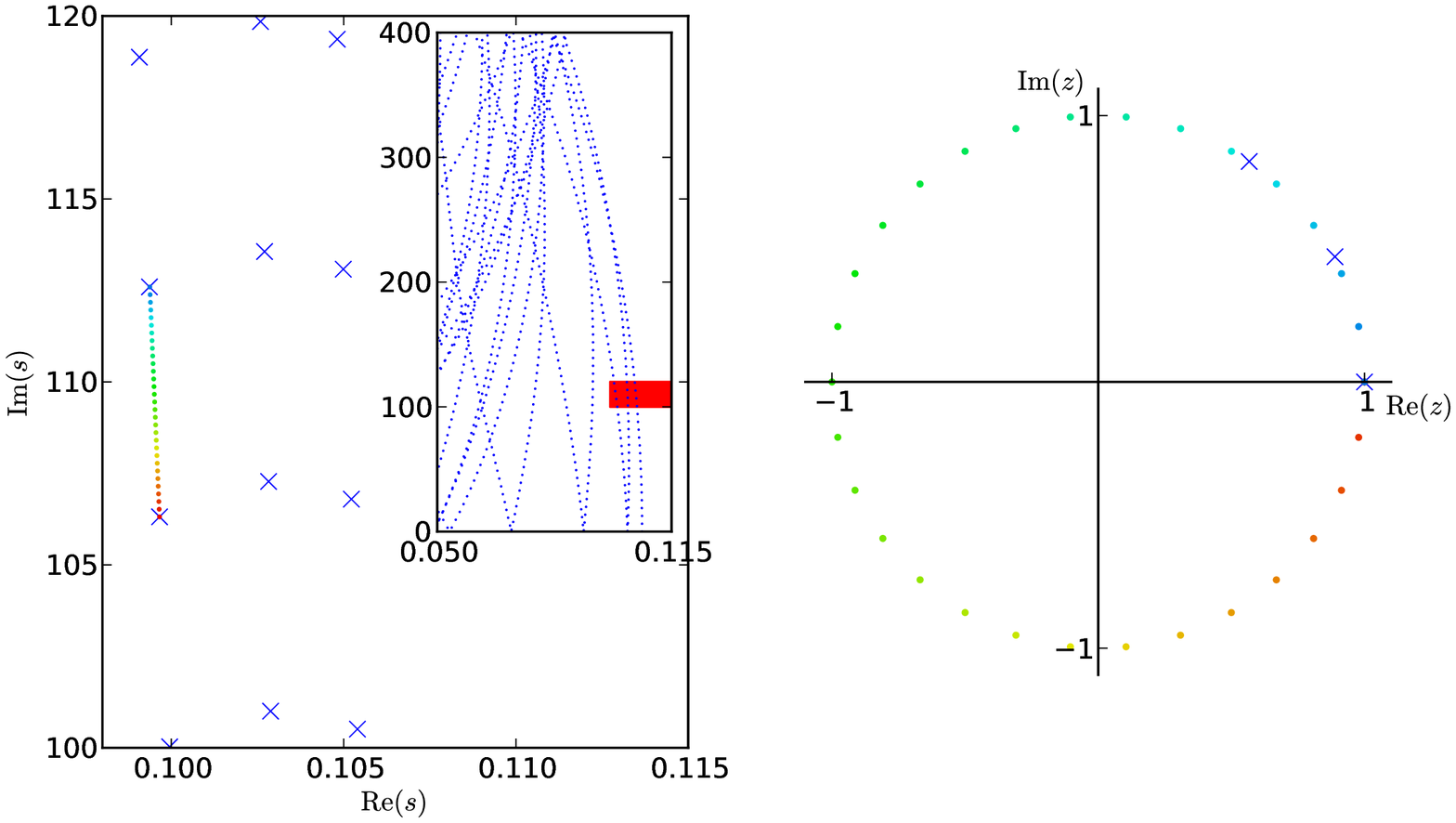}
\caption{Left: Plot of the resonances of $X_{12,13,14}$. The inset shows 
the chain structure of the resonances on a large domain in the complex 
plane. The main plot shows a zoom into one chain, such that the 
individual resonances become better distinguishable (blue crosses). For 
orientation the region of the zoom is indicated in the inset by a red rectangle. The 
colored dots between the two resonances $s_1=0.09966+106.31i$ and 
$s_2=0.09938+112.59i$ represent a discrete interpolation between the two 
resonances following the line suggested by the resonance chains. Right:
generalized spectrum for the order function as defined in
(\ref{eq:funnel_order_func}) with $n_1=12,n_2=13,n_3=14$ for spectral
values with $|z|>0.995$.
The blue crosses represent 
the spectrum $\sigma^{(\mathbf n)}_{s_1}$. By the colored dots we follow the 
evolution of one spectral value as the $s$-values 
evolve as indicated by the colored points in the right panel. We only follow the 
evolution of the spectral value starting at 1 for $s=s_1$ for more
clarity.}
\label{fig:rot_res_12_13_14}
\end{figure}

This approach of working with a generalized zeta function immediately allows us
to understand resonance chains for non-symmetric surfaces, as well. These chains have first
been observed by Borthwick, for example for the surface 
$X_{12,13,14}$ (cf.~\cite[Figure 7]{bor14}). A homogeneous return time 
would here require to cut one turn around the first funnel into 12 pieces,
one turn around the second funnel into 13 and a turn around the third funnel
into 14 pieces. We therefore define for $n_1=12, n_2=13, n_3=14$ the order
function
\begin{equation}\label{eq:funnel_order_func}
 \mathbf n:\Abb{P_X}{\mathbb N}{\gamma}{\sum\limits_{i=1}^{3}n_iw_i(\gamma)}
\end{equation}
where $w_i(\gamma)$ counts the windings of the geodesic around the $i$-th 
funnel. Figure~\ref{fig:rot_res_12_13_14} shows the generalized spectrum 
for this order function and it perfectly explains the resonance chains.
Again the spectral value turns once around the unit circle and 
creates the next resonance on the chains when the $s$ value moves along the 
chains. Note that the two other spectral values in 
$\sigma_{s_1}^{(\mathbf n)}$ (blue crosses in the left part of
\fref{fig:rot_res_12_13_14}) seem also to lie on the unit circle, which 
is not the case. Their absolute values are $1.0056$ and $1.0032$ and they 
belong to nearby chains, which are visible in the left part of 
\fref{fig:rot_res_12_13_14}. There are also further spectral values 
with absolute value slightly smaller then one. We however decided to 
not plot them in \fref{fig:rot_res_12_13_14} for more clarity and
as a reliable calculation of the generalized spectrum becomes more and more
complicated for small absolute values of $z$ as the order function 
$\mathbf n$ appears with high exponents (cf.~\ref{app:numeric}).

\subsection{The length spectrum and existence of resonance chains}
\label{sec:existence_of_chains}
From the examples in the previous section we have learned that the resonance
chains for symmetric as well as for asymmetric Schottky surfaces are best 
understood in terms of the generalized zeta function for a suitable choice 
of an order function. Up to now we have only presented two examples in
(\fref{fig:rot_res_GZ} and \fref{fig:rot_res_12_13_14}) together
with a good choice of an order function but we have not yet clearly worked 
out under which circumstances it is possible to construct a ``good'' order function. 
We will see that this question automatically leads us to a condition on the
observability of resonance chains. 

\begin{figure}
\centering
        \includegraphics[width=\textwidth]{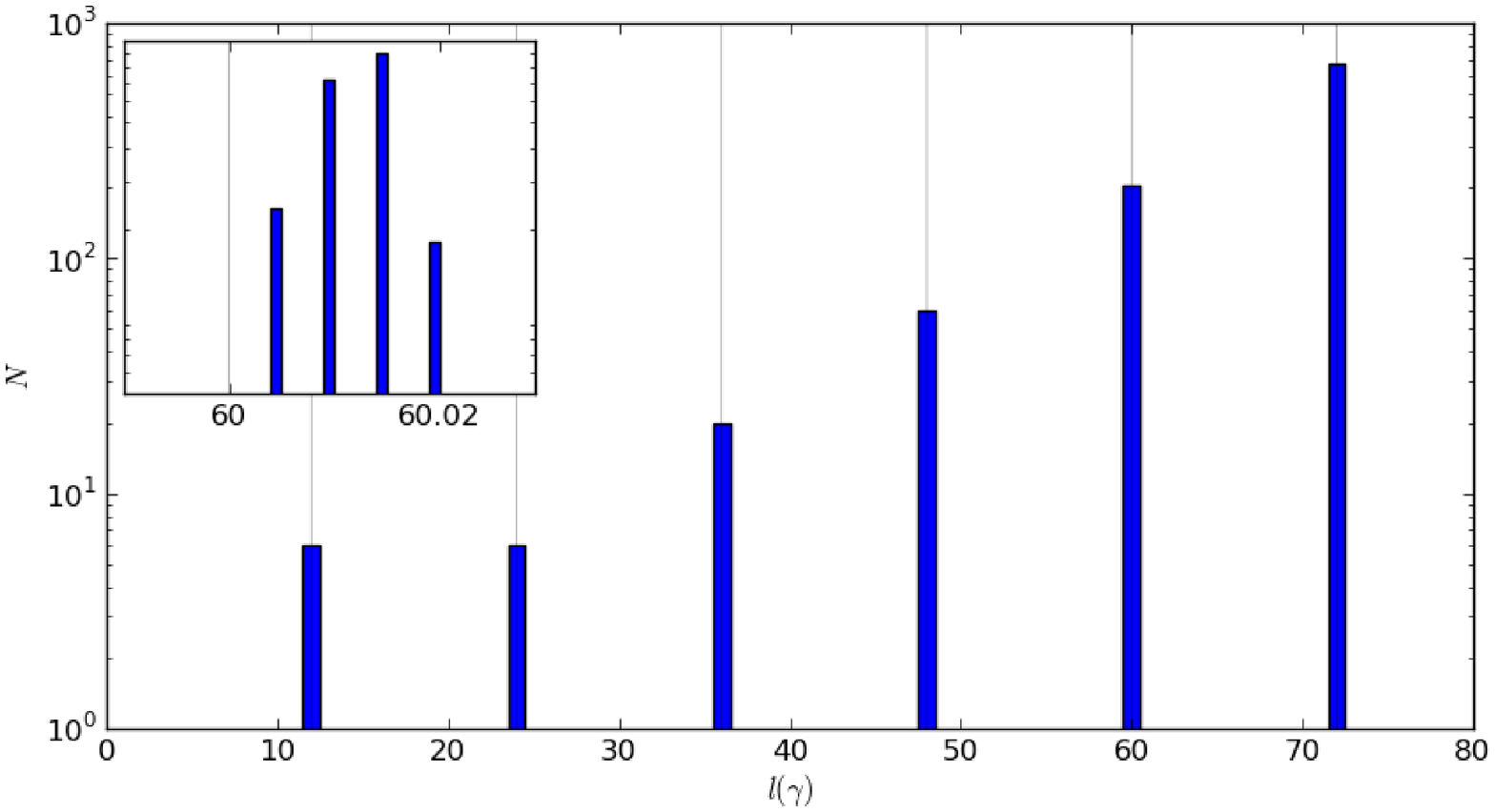}
\caption{Histogram of the primitive length spectrum for the surface 
$X_{12,12,12}$. The thin light lines indicate the integer multiples of
the base length $12$ and one observes that the lengths form clear clusters 
around these values. This clustering is however not perfect as can be seen 
in the inset where a histogram with a much smaller binsize resolves the peak 
at a length around 60. 
}
\label{fig:length_spec_12_12_12}
\end{figure}

\begin{figure}
\centering
        \includegraphics[width=\textwidth]{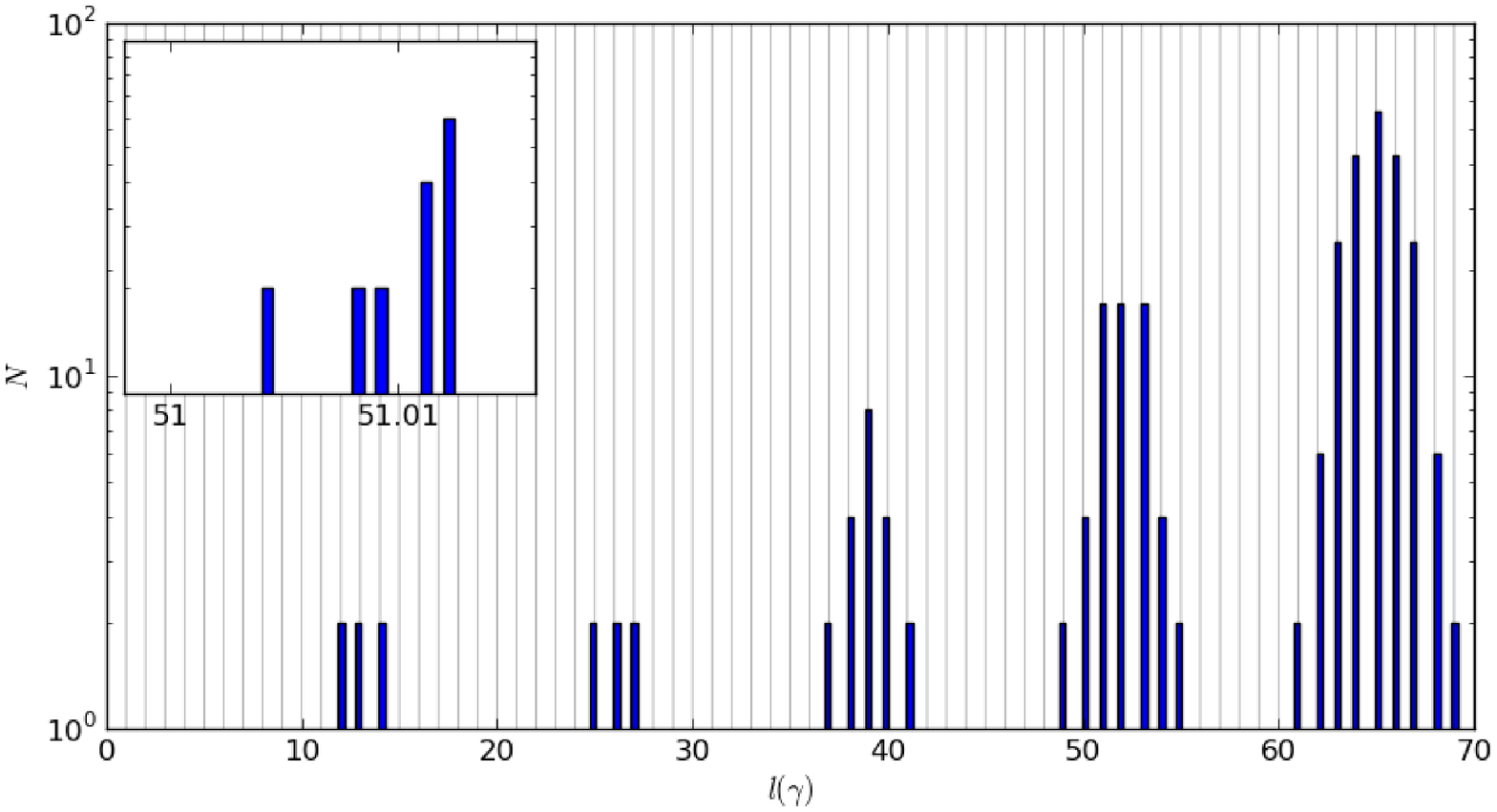}
\caption{Histogram of the primitive length spectrum for the surface 
$X_{12,13,14}$. The thin light lines indicate the integer multiples of
the base length $1$ and one observes that the lengths form clear clusters 
around these values. Again this clustering is not perfect as can be seen 
in the inset where a histogram with a much smaller binsize resolves the peak 
at a length around 51.}
\label{fig:length_spec_12_13_14}
\end{figure}

The primitive length spectrum, which contains the lengths of all 
primitive closed geodesics, repeated by multiplicity, can be seen 
for $X_{12,12,12}$ in \fref{fig:length_spec_12_12_12} and 
for $X_{12,13,14}$ in \fref{fig:length_spec_12_13_14}. We see that 
the length spectra of both surfaces form
a clear clustering on multiples of a base length $\ell$.

The big difference between these two cases is that the common base length to 
all clusters is $\ell=12$ for $X_{12,12,12}$ and $\ell=1$ for $X_{12,13,14}$. 
Analyzing the two order functions which we have used in 
these examples, we observe that they exactly respect this clustering: All
geodesics of one cluster are mapped to the same integer which is given 
by the multiple of the base length $\ell$ on which the cluster is located. 
All geodesics that are necessary to be taken into account for the calculation of
the resonances fulfill very well the approximation
\begin{equation}\label{eq:order_func_condition}
 l(\gamma)\approx \mathbf n(\gamma)\cdot\ell.
\end{equation}
If this condition is exactly 
fulfilled, then this will lead, similar to the case of a constant return time
in Section~\ref{sec:rot_res},
to the existence of straight resonance chains. In fact under this assumption
the generalized zeta function could be written as 
\[
 d_{\mathbf n}(s,z):= \prod_{\gamma\in P_X}\prod_{m=0}^\infty 
       \left(1-(ze^{-s\ell}e^{-m \ell})^{\mathbf n(\gamma)}\right),
\]
and if $s_0\in \tu{Res}_X$, i.e.~$d_{\mathbf n}(s_0,1)=0$ then for
all $b\in\R$ also $d_{\mathbf n}(s_0+ib,e^{ib\ell})=0$. This condition, 
however, directly implies that  $s_0+2\pi k i/\ell \in \tu{Res}_X$ 
for each $k\in \Z$, so all resonances would equidistribute on straight lines
with a distance of $2\pi/\ell$. Note that for the hyperbolic cylinder (\ref{eq:order_func_condition}) is exactly fulfilled and the resonance spectrum
is known to be ordered along straight lines \cite[Section 5.1]{bor07}. For non-elementary surfaces (\ref{eq:order_func_condition}) can, however, only hold
approximately, and one thus expects that these straight chains start to bend and that 
the spacing of the resonances along the chains becomes more irregular.

We therefore can formulate the following hypothesis:

\emph{If the primitive length spectrum of an infinite volume Riemannian surface (or an 
open Euclidean billiard) forms clusters at integer multiples of a base 
length, i.e.~if there is a length $\ell$ and an order function 
$\mathbf n:P_X\to \N$ such that (\ref{eq:order_func_condition})
holds, then one will observe resonance chains. These chains will be described
by the projection $\tu{Pr}_s (\mathcal C_{\mathbf n})$ of the analytic variety 
$\mathcal C_{\mathbf n}:=\{d_{\mathbf n}=0\}\cap\{|z|=1\}$ to the $s$-component and the 
resonances will be approximately spaced by a distance of $2\pi/\ell$
on these lines. The worse the approximation (\ref{eq:order_func_condition}) 
is fulfilled, the stronger the chains will bend and the more irregular 
the resonance distribution on the chains will become.}

Of course (\ref{eq:order_func_condition}) cannot hold on the whole 
length spectrum, but recall that for the calculation of the resonances
in a bounded subset of a complex plane, one only needs finitely many
geodesics. We thus suppose that the chains will be observable in this
bounded domain, if (\ref{eq:order_func_condition}) is fulfilled on the
geodesics which are necessary for their calculation.

\section{Further tests of the Hypothesis}\label{sec:num_test}
We already saw several examples in Section~\ref{sec:rot_res} 
and \ref{sec:gen_zeta} which support our hypothesis on the existence of resonance
chains. We will now test its validity against some more examples. First of all 
our condition should be able to explain the non-observability of resonance chains
in some systems which are structurally similar to systems which show resonance chains.

\begin{figure}
\centering
        \includegraphics[width=\textwidth]{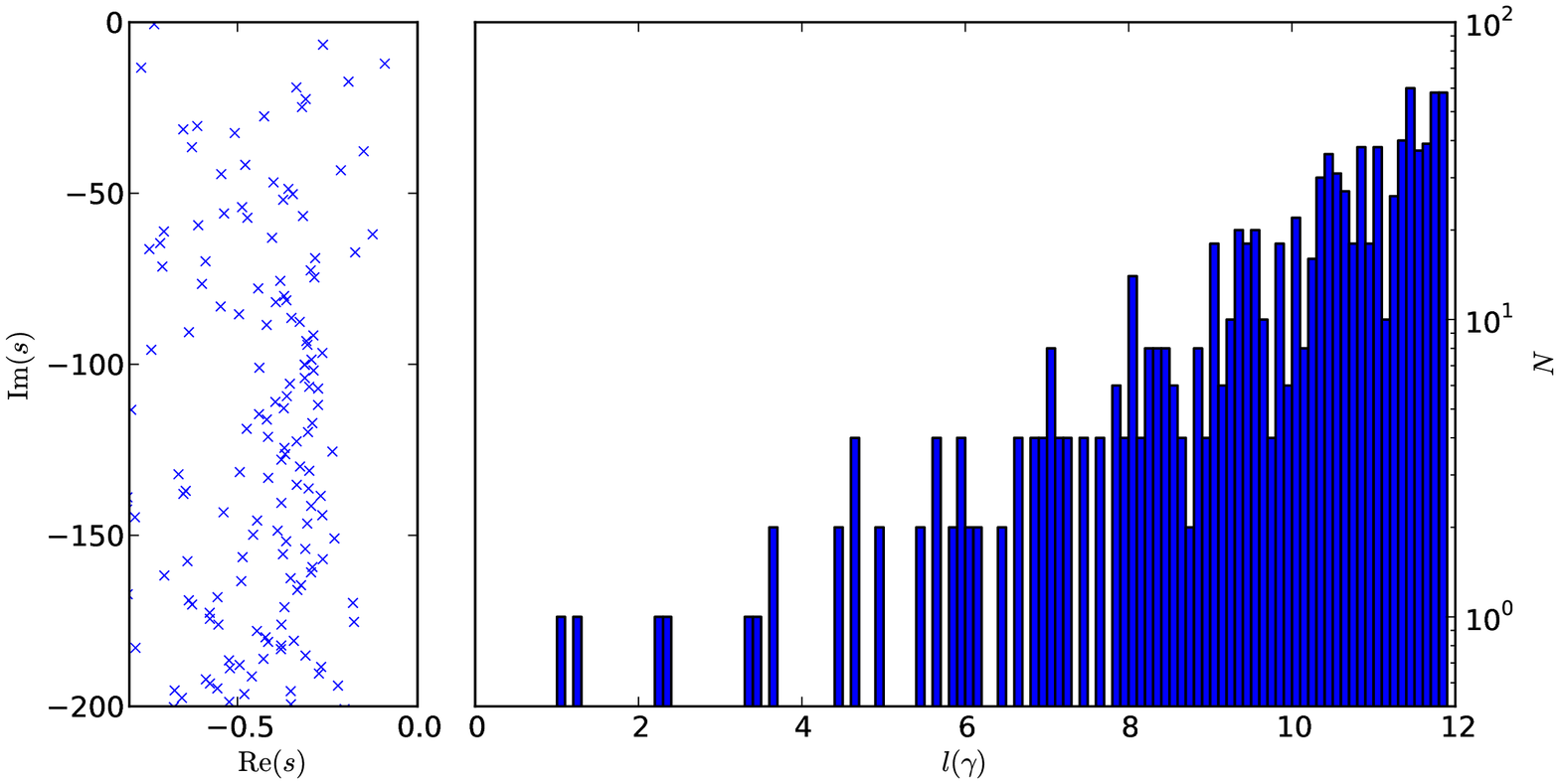}
        \includegraphics[width=\textwidth]{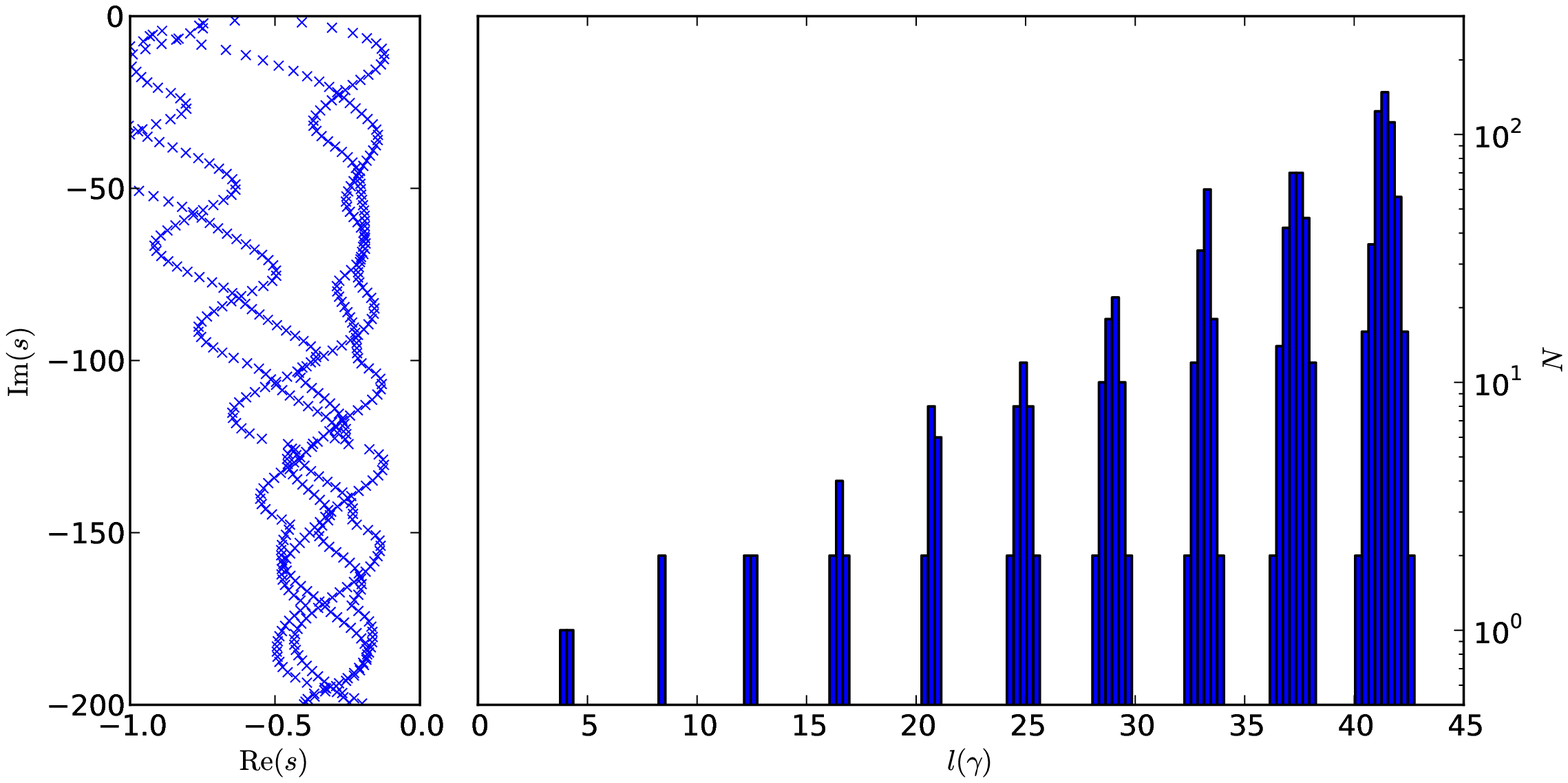}
\caption{Comparison of the resonance structure (left) with the 
primitive length spectrum (right) of the symmetry
reduced 3-disk system for $R/a=3$ (upper plot) and $R/a=6$ (lower plot).
While for $R/a=6$ one observes a clear clustering of the lengths and an obvious 
chain structure in the resonances, the lengths equidistribute very quickly and
their are no visible chains in the resonance spectrum.}
\label{fig:length_spec_3-disk}
\end{figure}

For example one observes that symmetric 3-disk billiards only show resonance chains
if the disks are sufficiently far away from each other. The 
resonance spectrum of the symmetric 3-disk billiard with $R/a=3$ 
shows no chain structure contrary to the more open case with $R/a=6$
(see \fref{fig:length_spec_3-disk}). This is
easily understood by looking at the primitive length spectrum. 
While for $R/a=6$ one sees a clear
clustering of the lengths, the length spectrum for $R/a=3$ equidistributes very 
quickly and it is impossible to choose an order function which fulfills 
(\ref{eq:order_func_condition}).
Our hypothesis also explains why symmetric 4-disk systems show in general
no chain structure (see e.g. FIG.~13 in \cite{gas92} for the Ruelle resonances 
of a 4-disk system) whereas the 3-dimensional 4-sphere billiard
shows them: If 4 disks are distributed on a square, the lengths between two 
neighboring disks and the length between diagonal disks are in general not 
multiples of each others but form two incommensurable base lengths, so it will 
not be possible to construct a good order function. For the 4-sphere billiard
where the spheres are placed on a regular tetrahedron, the distance between an 
arbitrary pair of spheres is equal and, if the spheres are sufficiently far away 
from each other, one expects a clustering of the length spectrum on multiples of
this base length. Exactly in this case of sufficiently open 4-sphere billiards
the chain structure has been observed (cf. Figure 5.2-5.9 in \cite{ebe10b}).

\begin{figure}
\centering
        \includegraphics[width=\textwidth]{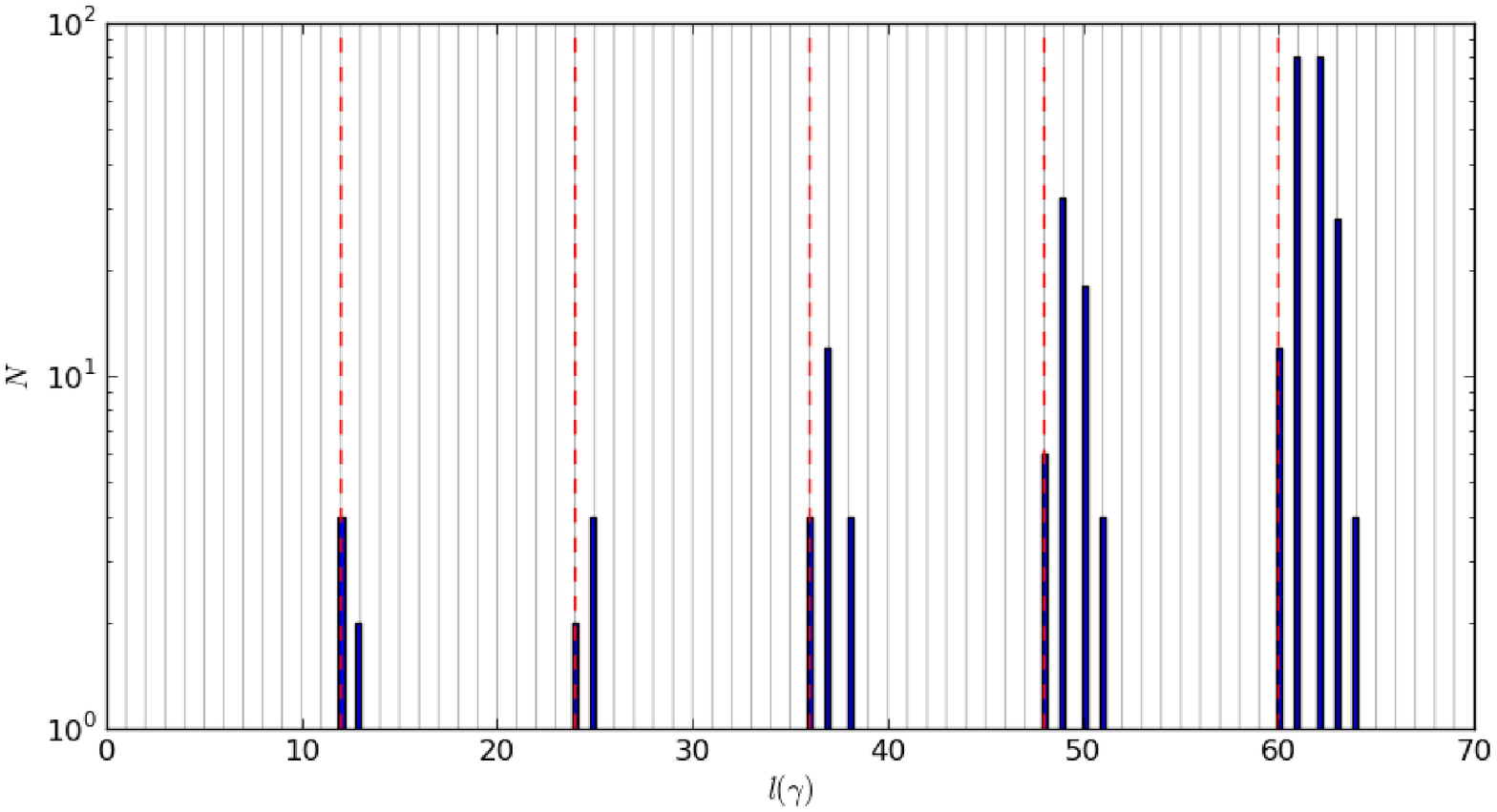}
\caption{Histogram of the primitive length spectrum for the Schottky surface 
$X_{12,12,13}$. The thin light lines indicate the integer multiples of
the base length $\ell=1$ and red dashed lines the multiples of a second base length
$\ell=12$. One can observe two types of clustering, one very coarse clustering
around the multiples of $\ell=12$ and a much finer clustering around the 
multiples of $\ell=1$.}
\label{fig:length_spec_12_12_13}
\end{figure}

\begin{figure}
\centering
        \includegraphics[width=\textwidth]{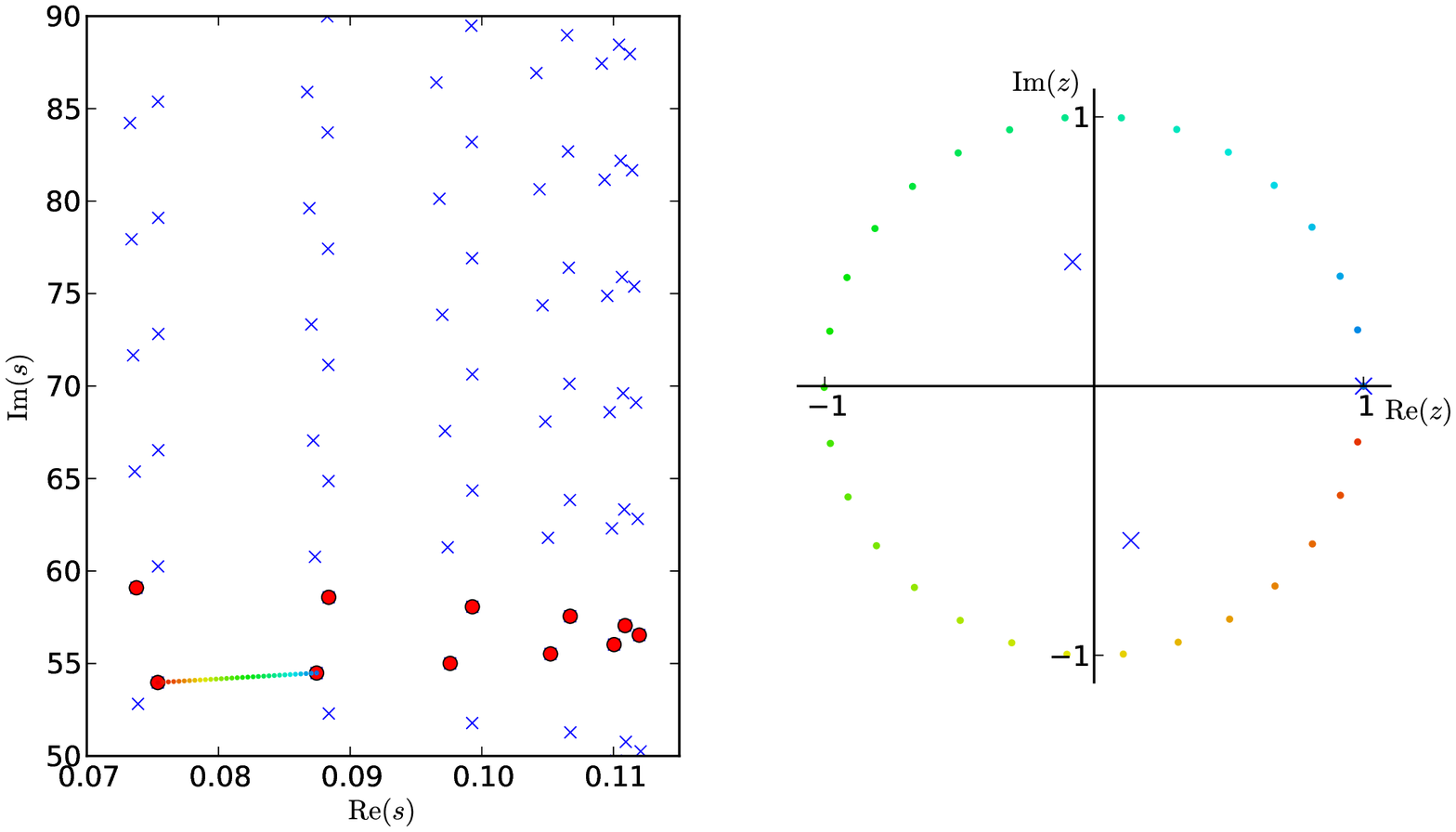}
\caption{
Left: The resonances of $X_{12,12,13}$ in the complex plane are marked
by crosses.  The red circles highlight the resonances on the chain, which 
belongs to the base length $\ell=12$. The 
colored dots between the two resonances $s_1=0.07539+53.97i$ and 
$s_2=0.08746+54.49i$ represent a discrete interpolation between the two 
resonances following the line suggested by the resonance chains. Right:
generalized spectrum for the order function as defined in
(\ref{eq:funnel_order_func})
with $n_1=n_2=n_3=1$. The blue crosses represent 
the spectrum $\sigma^{(\mathbf n)}_{s_1}$ for $|z|>0.2$. By the colored 
dots we follow the 
evolution of the spectral value which equals one while interpolating the 
$s$-values between $s_1$ and $s_2$ as indicated by the colored points on the
left.}
\label{fig:rot_res_12_12_13}
\end{figure}

\begin{figure}
\centering
        \includegraphics[width=\textwidth]{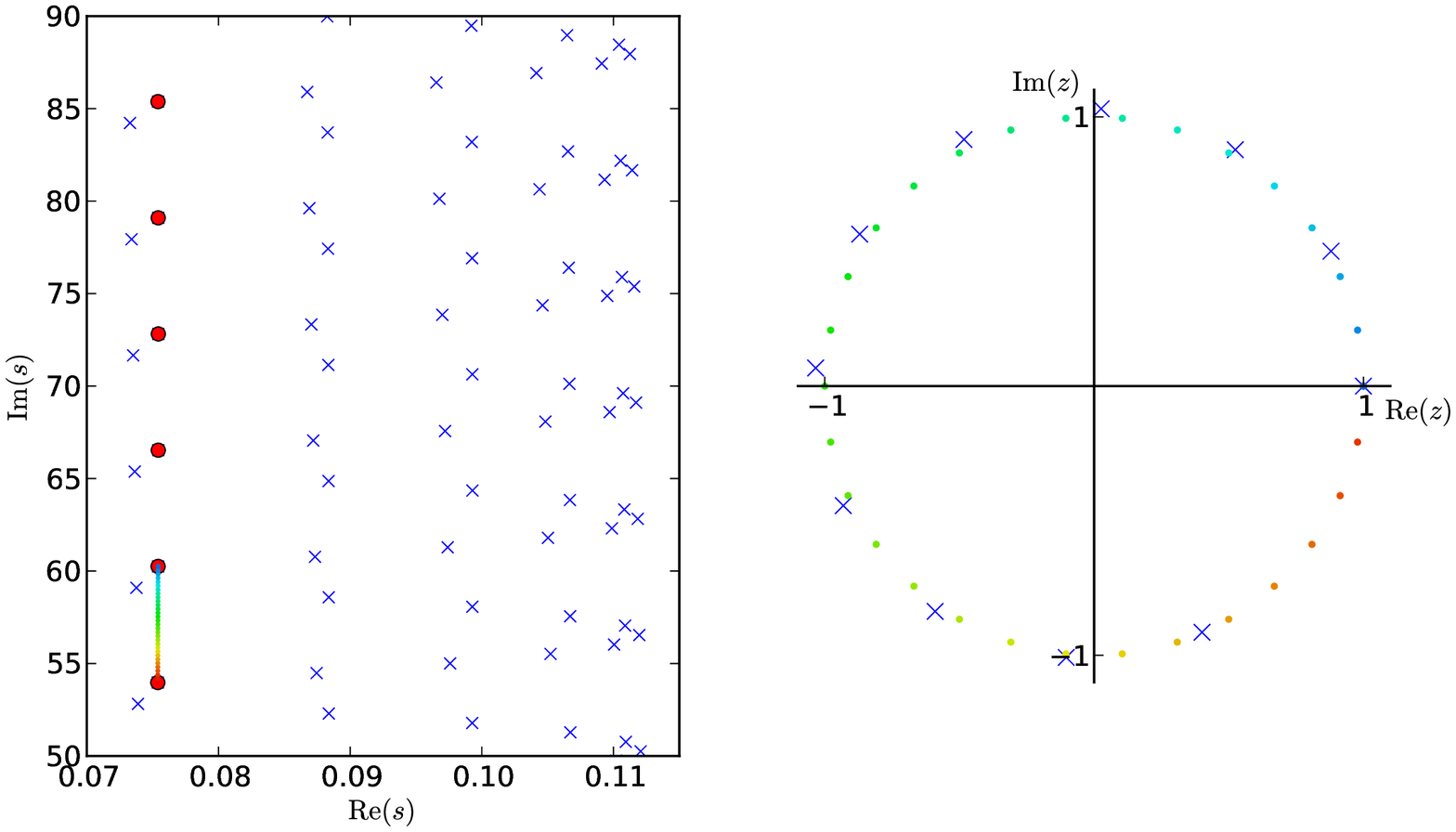}
\caption{Left: The resonances of $X_{12,12,13}$ in the complex plane are marked
by crosses.  The red circle highlight the resonances on the chain, which 
belongs to the base length $\ell=1$. The 
colored dots between the two resonances $s_1=0.07539+53.97i$ and 
$s_2=0.07542+60.25i$ represent a discrete interpolation between the two 
resonances following the line suggested by the resonance chains. Right:
generalized spectrum for the order function as defined in
(\ref{eq:funnel_order_func}) with $n_1=n_2=n_3=1$. The blue crosses represent 
the spectrum $\sigma^{(\mathbf n)}_{s_1}$ for $|z|>0.99$. By the colored 
dots we follow the 
evolution of the spectral value which equals one while interpolating the 
$s$-values between $s_1$ and $s_2$ as indicated by the colored points on the
left.}
\label{fig:rot_res_12_12_13_long}
\end{figure}

We will end this section by using the predictive power of our hypothesis in order 
to identify a new interesting system which shows the coexistence of two chains. 
As we only demand (\ref{eq:order_func_condition}) to hold approximately 
there should be examples for which there are different compatible base lengths 
and order functions, e.g.~the surface $X_{12,12,13}$. As can be seen in 
\fref{fig:length_spec_12_12_13}
the length spectrum can be either interpreted as forming clusters at multiples 
of the base length $\ell=12$ or at multiples of the base length $\ell=1$. 
The hypothesis thus predicts two different chain structures, one where the 
resonances are spaced by a distance of $2\pi/12$ and another one with a larger 
spacing of $2\pi$. As for $\ell=12$ the approximation 
(\ref{eq:order_func_condition}) holds much worse then for $\ell=1$ our hypothesis
predicts that the resonance chains belonging to $\ell=12$ are much more twisted
then the chains belonging to $\ell=1$. As figures~\ref{fig:rot_res_12_12_13} and
\ref{fig:rot_res_12_12_13_long} show, all these predictions are 
observed in the resonance structure. Choosing
the order function (\ref{eq:funnel_order_func}) with $n_1=n_2=n_3=1$ which
corresponds to the base length $\ell=12$ we observe strongly twisted resonance 
chains as indicated by the red circles in the left part of 
\fref{fig:rot_res_12_12_13}. The generalized spectrum $\sigma_s^{(\mathbf n)}$
of this order function creates this kind of chains by the rotating spectral 
values and if one connects neighboring resonances on this chain by the continuous
line which is suggested by the chain structure, the generalized spectral value
moves along the unit circle. This verifies that the chains are described by 
$\tu{Pr}_s(\mathcal C_{\mathbf n})$. Choosing, however, the order function 
(\ref{eq:funnel_order_func}) with $n_1=n_2=12, n_3=13$ which is the 
corresponding choice for the base length $\ell=1$ we obtain a generalized
zeta function which describes a second structure of resonance chains, on which 
the resonance distances are much larger and which are much straighter (see 
\fref{fig:rot_res_12_12_13_long}).


\section{Conclusion}
In this article we presented a unifying approach to the chain structure 
of quantum resonances, classical Ruelle resonances and geometric resonances. 
We showed at the example of 3-disk systems and Schottky surfaces that the
resonance chains can be understood in all three cases by means of a generalized 
zeta function $d_{\mathbf n}(s,z)$ which depends on the choice of an order function 
$\mathbf n:P\to\N$ on the set of primitive closed classical orbits $P$.
The central property of this generalized zeta function is that 
independent of the choice of the order function the resonances are given by the zeros of 
zeta function $Z(s)=d_{\mathbf n}(s,1)$. We showed that if the order function
is chosen such that 
\begin{equation}\label{eq:of_cond_concl}
l(\gamma)\approx \mathbf n(\gamma)\ell 
\end{equation}
for an arbitrary base length $\ell$ then the second complex variable allows to 
interpolate between the resonances of one chain. Furthermore we demonstrated 
that the continuous lines where the resonances are found on are given by the 
projection onto the $s$-component of the real analytic variety
\[
 \mathcal C_{\mathbf n}=\{d_{\mathbf n}(s,z) = 0\}\cap\{|z|=1\}.
\]
The existence of an order function that fulfills the condition (\ref{eq:of_cond_concl})
does of course depend on the structure of the 
length spectrum of the classical system. This led us to the hypothesis that 
the existence of resonance chains is directly linked to a clustering of the
classical length spectrum that allows us to choose an order function according
to (\ref{eq:of_cond_concl}). We finally validated this hypothesis by presenting
several examples of systems with and without resonance chains. In all cases 
the resonance chains were found to be linked to a clustering of the length 
spectrum. Furthermore we showed that this new understanding allows to construct
scattering systems with a customized resonance chain structure. This control 
of the resonance chains could be of importance in microdisk cavities 
where similar resonance chains have already been observed \cite{wie08} and
where a precise control of the resonance position in the complex plane is 
of great interest for their application as microdisk lasers. 

Additionally this understanding is crucial for fundamental questions
in quantum chaos. In chaotic systems the clear clusters in the length 
spectrum will only exist for short lengths. For larger lengths the clusters become 
broader and broader and will finally overlap. However, we required the 
condition (\ref{eq:of_cond_concl}) only to hold in the length range which is 
necessary for the calculation of the resonances in the studied frequency
regime. In other words for the existence of resonance chains in a given 
frequency range, condition (\ref{eq:of_cond_concl}) only has to hold for the lengths 
which can be resolved by the Planck cells at this frequency. If one still
observes resonance chains, then one has to keep in mind that one is not yet
at high enough frequencies to resolve the length spectrum in a regime, where 
the clusters dissolve. This could be especially important for the test of 
results that are obtained by arguing with phase cancellation such as the 
conjectures on the improved spectral gaps \cite{jak12}

Finally the study of resonance chains opens a variety of 
mathematical questions. We numerically observed in this article that the chain
structure becomes the clearer the better condition (\ref{eq:of_cond_concl}) is
fulfilled. By studying not just a single system but a whole family of scattering 
systems, where (\ref{eq:of_cond_concl}) is increasingly well fulfilled in 
a certain limit one can then try to prove the existence of resonance chains 
asymptotically in this limit. Furthermore one can try to give precise, simple 
formulas for the location
of these chains. For symmetric 3-disk systems this limit would correspond 
to large $R/a$-values and for Schottky surfaces to large funnel widths. For 
the latter case, first results into this direction will be presented in 
\cite{mathArticle}.

\ack
We are grateful to Ulrich Kuhl for many interesting stimulating discussions and
detailed feedback concerning this work.
Furthermore we thank Bruno Eckhardt, David Borthwick, Stéphane Nonnenmacher, Pablo Ramacher and Gabriel Rivière
for discussions and comments. This work was supported by the project ANR
2009-12 METHCHAOS. S.B. acknowledges financial support by DFG via the project FOR760 and
T.W. by the German National Academic Foundation.
\appendix
\section{Derivation of the product form for dynamical zeta function}\label{app:product_form}
In this appendix we will recall the derivation of the product form 
(\ref{eq:dynamical_zeta_product_form}) from the definition of the 
dynamical zeta function as it has been defined in (\ref{eq:dynamical_zeta_flat})
by the flat trace
\[
 d^\flat(z):= \exp\left(-\sum_{n>0}\frac{z^n}{n}\sum_{x\in \mathrm{Fix}(\phi^n)}\frac{V_n(x)}{|\det(1-(D\phi^n)(x))|} \right).
\]
For each fixed point $\phi^n(x)=x$ the Jacobian is a hyperbolic, symplectic
$2\times2$ matrix and it thus has two real eigenvalues $\Lambda_n$ and $1/{\Lambda_n}$ 
with $|\Lambda_n|>1$ and the determinant can be written as 
\begin{eqnarray}
 \frac{1}{|\det(1-(D\phi^n)(x)|}&=\frac{1}{|1-\Lambda_n||1-1/\Lambda_n|}&\nonumber\\
 &= |\Lambda_n|^{-1}\left(\sum\limits_{r\geq 0}\Lambda_n^{-r}\right)\left(\sum\limits_{s\geq 0}\Lambda_n^{-s}\right)& \nonumber\\
 &= |\Lambda_n|^{-1} \sum\limits_{k\geq 0}(k+1)\Lambda_n^{-k}&
\end{eqnarray}
where for the last equality we used that a positive integer $k$ can be 
written as $k+1$ different sums of two positive integers $r$ and $s$.

As a next step we can treat the double sum $\sum_{n>0}\sum_{x\in\mathrm{Fix}(\phi^n)}$ 
which is simply the sum over all fixed points of $\phi^n$ for arbitrary length $\phi$.
First we note that the iterated product of the potential function $V_n(x)$ as well as the instable 
eigenvalue $\Lambda_n$ do not depend on the choice of the fixed point in a 
fixed orbit $\{x,\phi(x),\ldots,\phi^{n-1}(x)\}$. Additionally we see that if an orbit of length 
$m\cdot n_p$ is a $m$-times iterate of a primitive orbit of length $n_p$, then 
the orbit contains $n_p$ fixed points and as well the iterated product as the 
instable eigenvalue are the $m$-th power of the values from the primitive orbit
\[
 V_{m\cdot n_p}(x) = (V_{n_p}(x))^m \tu{ and }\Lambda_{m\cdot n_p} = \Lambda_{n_p}^m.
\]
This allows us to write the double sum $\sum_{n>0}\sum_{x\in\mathrm{Fix}(\phi^n)}$ as
a double sum over all primitive periodic orbits and their repetitions $\sum_{p\in P}\sum_{m>0}$
and we obtain
\begin{eqnarray}
 d^\flat(z)&= \exp\left(-\sum_{k\geq0}(k+1)\sum_{n>0}\sum_{x\in \mathrm{Fix}(\phi^n)}
             \frac{z^n}{n}V_n(x) |\Lambda_n|^{-1}\Lambda_n^{-k}\right)&\nonumber\\
 &= \exp\left(-\sum_{k\geq0}(k+1)\sum_{p\in P}\sum_{m>0}
    \frac{z^{n_p\cdot m}}{m}(V_{n_p}(x))^m \left(|\Lambda_{n_p}|
     \Lambda_{n_p}^{-k}\right)^m\right)&\nonumber\\
 &=\prod_{k\geq0}\prod_{p\in P}\left[\exp\left(-\sum_{m>0}\frac{\left(z^{n_p}V_{n_p}(x)
   |\Lambda_{n_p}|^{-1}\Lambda_{n_p}^{-k}\right)^m}{m}\right)\right]^{k+1}&\nonumber \\
&= \prod_{k\geq0}\prod_{p\in P}\left[1-z^{n_p}\frac{V_{n_p}(x)}{|\Lambda_{n_p}|\Lambda_{n_p}^{k}}\right]^{k+1}&\nonumber
\end{eqnarray}
where we used the Taylor series of $\log(1-x)$ in the last equality.
\section{Numerical considerations for the calculation of the generalized 
spectrum $\sigma_s^{(\mathbf n)}$}\label{app:numeric}
In this appendix we will shortly recall how the resonance spectrum on 
Schottky surfaces is usually calculated (see \cite{jen02, gui04, bor14} for 
more details) and what modifications have to 
be performed for the calculation of the generalized zeta function and
the generalized spectrum. 

Usually the resonance spectrum 
is obtained by Taylor expanding (\ref{eq:BS_dyn_zeta_fixpoint_formula}) 
in $z$ around zero 
\[
 d_{BS}(s,z) = \sum_{k=0}^\infty z^k b_k(s).
\]
The Taylor coefficients are then explicitly given by \cite[Proposition 8]{jen02}
\begin{equation}\label{eq:b_k_Bowen_Series}
b_k(s)=\sum_{r=1}^k\sum_{(n_1,\ldots,n_r)\in P(k,r)} \frac{(-1)^r}{r!} \prod_{l=1}^r \frac{1}{n_l}\sum_{u\in\mathrm{Fix}(B^{n_l})}\frac{((B^{n_l})'(u))^{-s}}{1-(B^{-n_l})'(u)}
\end{equation}
where $P(k,r)$ are all $r$-partitions of $k$, i.e.~all $r$-tuples of integers 
whose sum equals $k$. The numerical task in order to calculate these coefficients
then consists in calculating the stabilities $(B^n)'(u)$
of sufficiently many fixed points. For a 3-funneled Schottky surface
$X_{l_1,l_2,l_3}$ this can efficiently be done using the generators of the 
corresponding Schottky group and a symbolic dynamic. 
For $X_{l_1,l_2,l_3}$ the generators can be written as
\[
 S_1=\left(\begin{array}{cc}
            \cosh(l_1/2)&\sinh(l_1/2)\\
            \sinh(l_1/2)&\cosh(l_1/2)
           \end{array}
\right),~~~ S_2=\left(\begin{array}{cc}
           \cosh(l_2/2)&a\sinh(l_2/2)\\
            a^{-1}\sinh(l_2/2)&\cosh(l_2/2)
           \end{array}
\right), 
\]
where the parameter $a$ is chosen such that $\Tr(S_1S_2^{-1})=-2\cosh(l_3/2)$ 
and as usual one writes $S_3=S_1^{-1}$ and $S_4=S_2^{-1}$. If we then define 
the set of words of length $n$ by
\[
 \mathcal W_n:=\Big\{w\in \{1,2,3,4\}^n,~|w_{j+1}-w_j|\neq r \tu{ for } j=1,\ldots,n-1\tu{ and } |w_1-w_m|\neq r \Big\}
\]
then we can define for each $w\in\mathcal W_n$ the hyperbolic isometry 
\[
 S_w:=S_{w_1}S_{w_2}\ldots S_{w_n}.
\]
One then can easily show from the definition of the Bowen-Series maps
\cite[Section 15.2]{bor07} that for each $u\in \tu{Fix}(B^n)$ there is 
exactly one $w \in \mathcal W_n$ and 
\[
 \Lambda(w):=2\cosh^{-1}\left(\frac{|\Tr(S_w)|}{2}\right)=(B^n)'(u)
\]
which allows an efficient calculation of (\ref{eq:b_k_Bowen_Series}) 
and thus of $d_{BS}(s,z)$.

For the calculation of $d_{\mathbf n}(s,z)$ one analogously has to Taylor-expand
the zeta function in $z$ around zero
\[
 d_{\mathbf n}(s,z) = \sum_{k=0}^\infty z^k b^{(\mathbf n)}_k(s).
\]
Then one has to calculate the new Taylor coefficients $b^{(\mathbf n)}_k(s)$.
As we want to use the word coding to calculate the lengths of the
geodesics or the stabilities $(B^n)'(u)$, respectively, we first have to transfer
the order function $\mathbf n:P_{X_{l_1,l_2,l_3}}\to\N$
to an order function on $\mathcal W=\bigcup_{n>0}\mathcal W_n$. 
This transformation can be obtained via the correspondence between 
closed geodesics on $X_{l_1,l_2,l_3}$ and fixed points of $B$ \cite[Proposition 15.5]{bor07}.
Using this identification one calculates that the order function 
as defined in (\ref{eq:funnel_order_func}) is then given by
\begin{equation}\label{eq:funnel_order_func_words}
 \mathbf n(w)=n(w_1,w_2)+n(w_2,w_3)+\ldots+n(w_n,w_1)
\end{equation}
where
\begin{eqnarray*}
 n(1,1)&=n(3,3)= n_1 ,~~ n(2,2)=n(4,4)= n_1 \\
 n(1,4)&=n(4,1)=n(2,3)=n(3,2)= n_3/2 \\
 n(1,2)&=n(2,1)= n(3,4)=n(4,3)=(n_1+n_2)/2. 
\end{eqnarray*}
The Bowen-Series order function $\mathbf n_{BS}$ can be translated much more easily.
From its definition in Section~\ref{sec:intro_Schottky} it immediately follows,
that $\mathbf n_{BS} (w) = n$ for all $w\in \mathcal W_n$.

Using the order function on the words as well as the identification of 
fixed points and words, equation (\ref{eq:BS_dyn_zeta_fixpoint_formula}) 
can be written as
\[
 d_{BS}(s,z)=\exp\left(-\sum_{j>0}\frac{1}{j}\sum_{w\in \mathcal W_j} z^{\mathbf n_{BS}(w)}\frac{\Lambda(w)^{-s}}{1-\Lambda(w)^{-1}} \right).
\]
Starting with (\ref{eq:gen_zeta}) and reversing the argumentation of \ref{app:product_form} the generalized zeta function
$d_{\mathbf n}(s,z)$ can be written as
\begin{equation}\label{eq:gen_zeta_exp_form}
 d_{\mathbf n}(s,z)=\exp\left(-\sum_{j>0}\frac{1}{j}\sum_{w\in \mathcal W_j} z^{\mathbf n(w)}\frac{\Lambda(w)^{-s}}{1-\Lambda(w)^{-1}} \right).
\end{equation}
An analogous calculation to those in \cite[Proposition 8]{jen02} then 
gives us the formula for the Taylor coefficients
\begin{equation}
 b^{(\mathbf n)}_k(s) = \sum_{r=1}^k\sum_{(n_1,\ldots,n_r)\in P(k,r)} \frac{(-1)^r}{r!} \prod_{l=1}^r \sum_{j>0}\frac{1}{j}\sum_{w\in\mathcal W_j,\tu{s.t. }\mathbf n(w)=n_l}\frac{\Lambda(w)^{-s}}{1-\Lambda(w)^{-1}}.
\end{equation}
The stabilities $\Lambda(w)$ can be calculated by the generators $S_i$ of 
the Schottky group as described above and in order to simplify the 
combinatorial task of building together the Taylor coefficients 
$b^{(\mathbf n)}_k(s)$ from these stabilities we can use the same 
recurrence trick as presented in \cite{gui04}. We therefore write
\[
 B^{(\mathbf n)}_{k,r}(s):=\sum_{(n_1,\ldots,n_r)\in P(k,r)} \frac{(-1)^r}{r!} \prod_{l=1}^r a^{(\mathbf n)}_{n_l}(s) 
\]
and
\[
a^{(\mathbf n)}_n(s):= \sum_{j>0}\frac{1}{j}\sum_{w\in\mathcal W_j,\tu{s.t. }\mathbf n(w)=n}\frac{\Lambda(w)^{-s}}{1-\Lambda^{-1}}.
\]
We can then use the recurrence relations
\[
 B^{(\mathbf n)}_{k,r}= \frac{1}{r}\sum\limits_{l=1}^{k-r+1}B^{(\mathbf n)}_{k-l,r-1}(s)a^{(\mathbf n)}_l(s).
\]
The only difference in calculating the generalized zeta function 
compared to the dynamical zeta function of the Bowen-Series maps
thus consists in the modified formula for the functions $a^{(\mathbf n)}_n(s)$.

Note that as $d_{\mathbf n}(s,z)$ is analytic the Taylor coefficients $b^{(\mathbf n)}_k(s)$
decay superexponentially in $k$ and theoretically one can compute $d_{\mathbf n}(s,z)$ for 
arbitrary $(s,z)$. Practically, however, the convergence becomes worse and worse
the smaller $\tu{Re}(s)$ and the bigger $|z|$ become. Especially for order functions
(\ref{eq:funnel_order_func}) with larger values of $n_1,n_2,n_3$ as they appeared 
for example for the surfaces $X_{12,13,14}$ the convergence depends very strongly
on $|z|$. This implies that a reliable calculation of $d_{\mathbf n}$ is only 
possible for $z$ values which are slightly greater then $1$. In terms of the 
generalized spectrum this implies that the numerical calculation is only
possible for $z$ slightly smaller then $1$. As the interesting region for the 
understanding of the resonance chains is, however, the unit circle, this 
is no severe problem for the numerical investigations presented in this article.
\section{Topological pressure and analyticity of $d^{(b)}(s,z)$}\label{app:top_pres}
The topological pressure of a 3-disk system is a function $P(\beta)$ of
a positive parameter $\beta>0$ that describes the convergence behavior of certain 
zeta functions. Writing 
\[
V^\beta_s(x):=|\Lambda(x)|^{1-\beta}e^{-s\tau(x)} 
\]
the topological pressure of the 3-disk system can be defined \cite{gas89b} 
to be the real number $P(\beta)$ such that
\[
d_\beta(s):=\det(1-\mathcal L_{V_s^\beta}) =\exp\left(-\sum\limits_{n>0}\frac{1}{n} \sum\limits_{x\in\tu{Fix}(\phi^n)} \frac{(V_s^\beta)_n(x)}{|\det(1-(D\phi^n)(x)|}\right) 
\]
is absolutely convergent for $\tu{Re}(s)>P(\beta)$. This directly implies that 
$d^{(a)}$ has no zeros for $\tu{Re}(s)>P(1/2)$ and that $d^{(b)}(s)$ has
no zeros for $\tu{Re}(s) > P(3/2)$. For a symmetric 3-disk system
with $R/a=6$ this value of the topological pressure is given by
\[
 P(3/2)=-0.699.
\]
Consequently all zeros of the Gutzwiller-Voros zeta function $Z_{GV}(s,1)$
with $\tu{Re}(s)>-0.699$ are automatically zeros of the Fredholm determinant
$d_{(a)}(s,1)$ and can thus be interpreted by the spectrum of the transfer 
operator $\mathcal L_{V_s^{(a)}}$. Note that all the numerical investigations
presented in this article are within this $s$-range.

\end{document}